\documentclass[fleqn,usenatbib]{mnras}

\usepackage{newtxtext,newtxmath}

\usepackage[T1]{fontenc}

\DeclareRobustCommand{\VAN}[3]{#2}
\let\VANthebibliography\thebibliography
\def\thebibliography{\DeclareRobustCommand{\VAN}[3]{##3}\VANthebibliography}

\usepackage{graphicx}	
\usepackage{amsmath}	
\usepackage{amssymb}	
\usepackage{bm}
\usepackage{caption}
\usepackage{multicol}
\usepackage{hyperref}
\usepackage{booktabs}
\usepackage{ulem}

\def\eqref#1{equation~(\ref{#1})}

\newcommand{\fracL}[2]{\frac{\displaystyle #1}{\displaystyle #2}}

\newcommand {\Gyr}{\,{\rm Gyr}}

\newcommand {\kpc}{\,{\rm kpc}}
\newcommand {\kms}{\,{\rm km}\,{\rm s}^{-1}}
\newcommand {\kmskms}{\,{\rm km}^{2}\,{\rm s}^{-2}}
\newcommand {\kpckms}{\,{\rm kpc}\,{\rm km}\,{\rm s}^{-1}}
\newcommand {\kmskpc}{\,{\rm km}\,{\rm s}^{-1}\,{\rm kpc}^{-1}}

\newcommand {\marcs}{\,{\rm mas}}
\newcommand {\marcsyr}{{\rm \, mas \, yr^{-1}}}
\newcommand {\dex}{\,{\rm dex}}
\newcommand {\dexkpc}{\,{\rm dex/kpc}}

\newcommand {\zsun}{\,{\rm z}_\odot}

\newcommand {\vRsun}{v_{R\odot}}
\newcommand {\vphisun}{v_{\varphi\odot}}
\newcommand {\vzsun}{v_{{\rm z}\odot}}

\newcommand {\Vsun}{{V_{\!\odot}}}

\newcommand {\sigpar}{{\sigma_{\rm p}}}

\newcommand {\vlos}{v_{\rm los}}

\newcommand {\gb}{{b}}

\newcommand {\mul}{\mu_l}
\newcommand {\mub}{\mu_b}

\newcommand {\vc}{v_{\rm c}}

\newcommand {\drm}{\mathrm{d}}
\newcommand {\vx}{{\bm x}}
\newcommand {\vvel}{{\bm v}}

\newcommand {\Rg}{R_{\rm g}}

\newcommand {\vR}{v_R}

\newcommand {\vphi}{v_\varphi}

\newcommand {\vz}{v_z}

\newcommand {\RCR}{R_{\rm CR}}

\newcommand {\Lz}{L_z}

\newcommand {\phib}{\varphi_{\rm b}}

\newcommand {\vJ}{{\bm J}}

\newcommand {\JR}{J_R}

\newcommand {\Jz}{J_z}

\newcommand {\Jphi}{J_\varphi}

\newcommand {\Jf}{J_{\rm f}}

\newcommand {\Js}{J_{\rm s}}

\newcommand {\Jsres}{J_{\rm s, res}}

\newcommand {\Jl}{J_\ell}
\newcommand {\Jlsep}{J_{\ell,{\rm sep}}}

\newcommand {\Jln}{\hat{J}_\ell}

\newcommand {\Jlnmin}{\hat{J}_{\ell,{\rm min}}}

\newcommand {\vtheta}{{\bm \theta}}

\newcommand {\thetaR}{\theta_R}
\newcommand {\thetaphi}{\theta_\varphi}

\newcommand {\thetas}{\theta_{\rm s}}
\newcommand {\dthetas}{\dot{\theta}_{\rm s}}
\newcommand {\ddthetas}{\ddot{\theta}_{\rm s}}

\newcommand {\thetasep}{\theta_{\rm sep}}
\newcommand {\thetassep}{\theta_{\rm s,sep}}

\newcommand {\Omegap}{\Omega_{\rm p}}

\newcommand {\Omegapi}{\Omega_{\rm p1}}
\newcommand {\dOmegap}{\dot{\Omega}_{\rm p}}

\newcommand {\OmegaR}{\Omega_R}

\newcommand {\Omegaphi}{\Omega_\varphi}

\newcommand {\vOmega}{\bm \Omega}

\newcommand {\NR}{N_R}

\newcommand {\Nphi}{N_\varphi}

\newcommand {\Ep}{E_{\rm p}}

\newcommand {\FeH}{[{\rm Fe/H}]}
\newcommand {\MH}{[{\rm M/H}]}

\newcommand {\meanMH}{\langle[{\rm M/H}]\rangle}

\newcommand {\mzi}{\mu_{z_i}}
\newcommand {\szi}{\sigma_{z_i}}

\newcommand {\pd}{{\partial}}

\title[]{Tree-ring structure of Galactic bar resonance}

\author[R. Chiba et al.]{
Rimpei Chiba$^{1}$\thanks{E-mail: rimpei.chiba@physics.ox.ac.uk}
and Ralph Sch{\"o}nrich$^{2}$
\\
$^{1}$University of Oxford, Rudolf Peierls Centre for Theoretical Physics, OX1 3PU Oxford, UK\\
$^{2}$Mullard Space Science Laboratory, University College London, Holmbury St. Mary, Dorking, Surrey, RH5 6NT, UK
}

\date{Accepted XXX. Received YYY; in original form ZZZ}

\pubyear{2021}

\begin{document}
\label{firstpage}
\pagerange{\pageref{firstpage}--\pageref{lastpage}}
\maketitle

\begin{abstract}

Galaxy models have long predicted that galactic bars slow down by losing angular momentum to their postulated dark haloes. When the bar slows down, resonance sweeps radially outwards through the galactic disc while growing in volume, thereby sequentially capturing new stars at its surface/separatrix. Since trapped stars conserve their action of libration, which measures the relative distance to the resonance centre, the order of capturing is preserved: the surface of the resonance is dominated by stars captured recently at large radius, while the core of the resonance is occupied by stars trapped early at small radius. The slow-down of the bar thus results in a rising mean metallicity of trapped stars from the surface towards the centre of the resonance as the Galaxy's metallicity declines towards large radii. This argument, when applied to Solar neighbourhood stars, allows a novel precision measurement of the bar's current pattern speed $\Omegap = 35.5 \pm 0.8 \kmskpc$, placing the corotation radius at $\RCR = 6.6 \pm 0.2 \kpc$. With this pattern speed, the corotation resonance precisely fits the Hercules stream in agreement with kinematics. Beyond corroborating the slow bar theory, this measurement manifests the deceleration of the bar of more than $24\%$ since its formation and thus the angular momentum transfer to the dark halo by dynamical friction. The measurement therefore supports the existence of a standard dark-matter halo rather than alternative models of gravity.

\end{abstract}




\section{Introduction}
\label{sec:introduction}

Many spiral galaxies, including our Milky Way, host a rotating bar at their centres. It has been predicted analytically \citep{Tremaine1984Dynamical,weinberg1985evolution,Weinberg2004Timedependent,Weinberg2007BarHaloInteraction} and confirmed by simulations \citep[e.g.][]{hernquist1992bar,debattista2000constraints,athanassoula2003determines,Martinez2006Evolution,Sellwood2008BarHaloIII} that these galactic bars experience dynamical friction against the postulated dark matter in the Galactic halo. This transfer of energy and angular momentum leads to less cuspy halo profiles \citep{Weinberg2002BarDriven}, but also slows the bar and allows it to grow \citep{athanassoula1992existence}. 

The bar affects the stellar disc most effectively at resonances where the bar's pattern speed is in commensurable relation with the orbital frequencies of stars. A well-known consequence of the non-linear response near resonances is that orbits there could become trapped. The trapped orbits occupy a distinct volume of phase space bounded by the separatrix. When the bar decelerates, these resonant islands sweep radially outwards through the stellar disc \citep{Chiba2020ResonanceSweeping}. As resonances migrate outwards, their phase-space volume generally grows, as we will demonstrate, so surrounding stars are sequentially captured into the resonance from the separatrix. Once in resonance, these trapped stars are dragged along with the resonance while approximately conserving their action of libration which characterizes the distance to the resonance centre. As a result, stars trapped at the early epoch of bar formation remain confined to the core of the resonance, while newly trapped stars fill in the phase space opened up by the expanding separatrix; the bar resonance evolves analogously to the rings of a growing tree where new layers of cells form at the bark and continuously record the climate condition of that time. Since stars trapped later originate from larger radii with typically lower metallicity, the deceleration history of the bar is imprinted on the metallicity distribution inside the resonance.

The tree-ring structure of the bar resonance has a bi-directional application: Once the current pattern speed of the bar and thus the location of the resonance is known, we can read out the history of bar evolution by looking at the variation of stellar metallicity with distance to the resonance centre. Reversely, we can pin down the current bar pattern speed by demanding the metallicity inside the resonance to rise from the surface towards the core as expected from the deceleration of the bar. 

Recent studies consistently provide evidence for a slow bar\footnote{Here `slow' vs. `fast' does not refer to the ratio between corotation radius and bar length as in studies of external galaxies, but exclusively describes slow pattern speed $\Omegap \lesssim 40 \kmskpc$ instead of the formerly favoured fast models with $\Omegap \gtrsim 50 \kmskpc$. See section \ref{sec:relatve_pattern_speed} for a discussion.}, 
although the precise bar pattern speed has not yet clinched: \cite{Portail2017Dynamical} derived $\Omegap = 39 \pm 3.5 \kpckms$ by adopting their N-body model to density and kinematics of the red clump giants. Their model was used by \cite{perez2017revisiting} to explain the Hercules stream in the Solar neighbourhood with the bar's corotation resonance and was further examined by \cite{Monari2019signatures} who showed that not only the Hercules but many of the prominent structures in local velocity and action space are well reproduced by the respective resonances of the bar, although they note that a slightly lower pattern speed (or a lower Solar azimuthal velocity) will yield better agreement with the data. \cite{Clarke2019Milky} showed that the integrated on-sky maps of the mean longitudinal proper motion is consistent with models with $\Omegap = 37.5 \kmskpc$. \cite{Sanders2019pattern} used the continuity equation and derived $\Omegap = 41 \pm 3\kmskpc$ from proper motions of stars in the near side of the bar and $\Omegap = 31 \pm 1\kmskpc$ when data on the far side is also considered. A similar conclusion is reported by \cite{bovy2019life} where they determined the pattern speed purely from kinematic data. \cite{Binney2020Trapped} applied Jeans’ theorem to trapped orbits in the Solar neighbourhood and showed that its violation is minimized at $\Omegap = 36 \pm 1 {\rm Gyr}^{-1} = 35.2 \pm 1 \kmskpc$.

Given the above uncertainties, in this paper, we take the second approach: constraining the bar pattern speed from the metallicity ordering inside the resonance. Using photometric metallicity derived from \textit{Gaia} DR2, we show that demanding a monotonic trend in metallicity inside the corotation resonance of the bar narrowly constrains the current pattern speed to $\Omegap = 35.5 \pm 0.8 \kmskpc$ where the corotation resonance perfectly fits the Hercules stream in agreement with kinematic models \citep[e.g.][]{perez2017revisiting,Monari2019signatures,DOnghia2020Trojans,Binney2020Trapped,Chiba2020ResonanceSweeping}.

In the following section we provide a brief description on resonant dynamics which underlies our analysis. In section \ref{sec:method}, the method of estimating metallicity from \textit{Gaia} photometry is described and tested with open clusters which the metallicity is measured independently from spectroscopy. In section \ref{sec:results}, we present the mean metallicity map in the local velocity/action plane and subsequently constrain the bar pattern speed by evaluating the likelihood of monotonic increase towards the resonance centre. Systematic errors due to sample selection and uncertainties in model parameters are reported. Section \ref{sec:conclusion} sums up.

\section{Theoretical background}
\label{sec:theory}

\subsection{Adiabatic invariants of resonantly trapped orbits}
\label{subsec:adiabatic_invariants}

Just like Jupiter's Trojans and Greeks, stars in the Galactic disc can be trapped in corotation resonance with the bar. In the rotating frame of the bar, their orbits slowly circulate around the stable Lagrange points $L_{4,5}$ along the bar's minor axis. To describe orbital motions, galactic dynamics relies on `actions' $\vJ$, which are conserved under adiabatic (slow) changes of the potential, and their canonically conjugate `angle' variables $\vtheta$, which evolve linearly with time at constant rates $\vOmega$. In axisymmetric potentials, there are three actions. One commonly uses $(\JR, \Jphi, \Jz)$: the radial action $\JR$ encodes the size of radial oscillations (and thus orbital eccentricity), the azimuthal action $\Jphi$ is the angular momentum component along the rotation axis, and the vertical action $\Jz$ quantifies oscillations perpendicular to the disc plane. The kinematic substructures found in the Solar neighbourhood weakly depends on $\Jz$ \citep{friske2019more}, so in the following we describe the Galactic disc by a two-dimensional model. In the presence of a non-axisymmetric bar, $\JR$ and $\Jphi$ are no longer conserved but fluctuate with amplitude increasing towards resonances, i.e. $ \NR \OmegaR + \Nphi (\Omegaphi - \Omegap) = 0$, where stars can be trapped. The dynamics near resonance exhibits slow motion around the resonance and is dealt by secular perturbation theory where the Hamiltonian is averaged over the fast motions \citep{lichtenberg1992regular}. The resulting Hamiltonian takes the form similar to that of a pendulum and the slow dynamics of trapped orbits is described in the \textit{slow} angle-action plane $(\thetas, \Js)$ with one degree of freedom, where
\begin{equation}
  \thetas \equiv \NR \thetaR + \Nphi \left(\thetaphi - \int dt~\Omegap \right) ~~ \text{and} ~~ \Js \equiv \frac{\Jphi}{\Nphi}.
  \label{eq:slow_angle}
\end{equation}
The phase-space in $(\thetas, \Js)$ is split by the separatrix into regimes of libration (trapped orbits), and circulation (non-trapped orbits). Trapped orbits have two approximate constants of motions: the fast action 
$\Jf \equiv \JR - (\NR/\Nphi) \Jphi$ ($\JR$ in the case of co-rotation resonance), and the action of libration
\begin{equation}
  \Jl \equiv \frac{1}{2\pi}\oint d\thetas~\Js(\thetas),
  \label{eq:libration_action}
\end{equation}
which characterizes the amplitude of motion around the resonance and takes maximal value at the separatrix. The conservation of $\Jl$ is subject to the adiabaticity condition that the libration period, which diverges to infinity at the separatrix, is significantly smaller than the migration timescale of the resonance. $\Jl$ is thus not conserved near the separatrix, which allows orbits to enter or leave the resonance there.

\subsection{Increase in phase-space volume of resonance}
\label{subsec:increase_in_resonant_volume}

\begin{figure}
  \begin{center}
    \includegraphics[width=8.5cm]{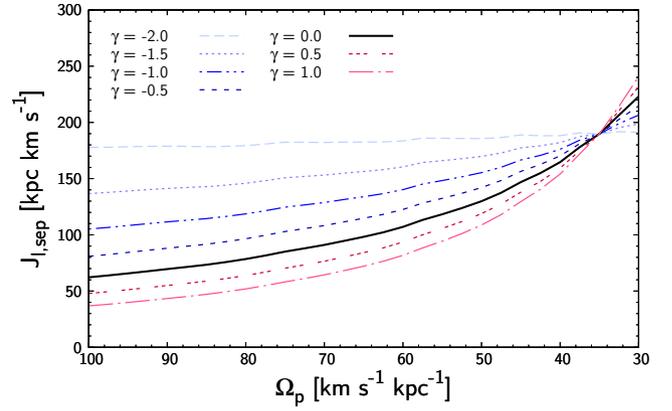}
    \caption{Phase-space volume of the bar's corotation resonance characterized by the libration action at the separatrix as a function of the decreasing pattern speed. The fast action is fixed to $\Jf = 50 \kpckms$. Models with different bar growth/decay rate $\gamma$ are plotted. All models take the same current bar amplitude $A=0.02$ at the current pattern speed $\Omegap = 35 \kmskpc$. The resonance volume generally grows unless the bar amplitude decay rapidly $(\gamma \lesssim -2)$.}
    \label{fig:wp_Jlsep}
  \end{center}
\end{figure}

The phase-space volume occupied by each resonance is described by the libration action evaluated at the separatrix $\Jlsep$ (equation \ref{eq:Jlsep}). Figure~\ref{fig:wp_Jlsep} shows $\Jlsep$ of the CR at $\Jf = 50 \kpckms$ as a function of the decreasing pattern speed. We present various bar models (Appendix \ref{sec:app_model}) with their strength varying according to 
\begin{equation}
  A(t) = A_1 \left[\frac{\Omegap(t)}{\Omegapi}\right]^{-~\gamma},
  \label{eq:bar_amplitude_growth}
\end{equation}
such that it takes the current amplitude $A_1=0.02$ at the current pattern speed $\Omegapi = 35 \kmskpc$. The functional form of $\Omegap(t)$ is irrelevant here but, for instance, with $\Omegap(t) \propto t^{-1}$ as assumed in our simulations, the bar strength scales as $A(t) \propto t^{\gamma}$. We show all cases where $A$ is constant ($\gamma=0$, black), decreases ($\gamma<0$, blue), or increases ($\gamma>0$, red). The phase-space volume of the resonance $\Jlsep$ increases monotonically unless the bar weakens significantly $(\gamma \lesssim -2)$ while slowing down. N-body simulations show that deceleration of the bar is typically accompanied by an increase in bar amplitude \citep[e.g][]{debattista2000constraints,Martinez2006Evolution,Ghafourian2020Modified} except at the early buckling phase, so we expect the resonance to grow and thus continuously sweep up stars as it moves outwards. It can be trivially shown that $\Jlsep$ scales as $\sqrt{A} \vc^2 / \Omegap$ in the epicycle limit (Appendix \ref{sec:app_librationaction}) which explains why $\Jlsep$ is approximately constant when $\gamma = -2$.

\subsection{Tree-ring structure of resonance}
\label{subsec:tree_ring_structure_of_resonance}

\begin{figure}
  \begin{center}
    \includegraphics[width=8.3cm]{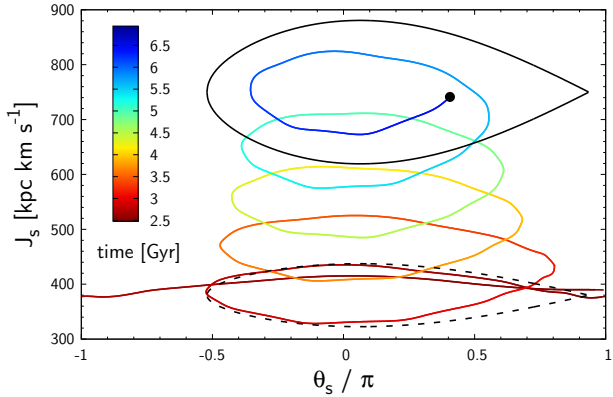}
    \caption{Trajectory of a test particle trapped and dragged by the corotation resonance of a decelerating bar (Appendix \ref{sec:app_model}) where the colour indicates time. The black curves represents the separatrix at the time of capture (dashed) and at the end of the simulation (solid). The star streams along the separatrix at the moment of capture but is then detached from the separatrix as the volume of resonance grows.}
    \label{fig:single_slowAA}
  \end{center}
\end{figure}

When a resonance grows in volume, a fraction of stars outside the resonance that are passed by the separatrix may be captured into the resonance depending on their angular phase (see Appendix \ref{app:capture_probability} for the angle-averaged capture probability). Figure~\ref{fig:single_slowAA} shows a typical stellar trajectory captured into the sweeping resonance. The star initially circulates at $\Js \sim 400 \kpckms$ above the resonance, while the separatrix approaches from below. The star then passes the separatrix, becomes trapped in the resonant region and gets dragged towards larger $\Js$ (and thus towards larger angular momentum $\Jphi$) while keeping its $\Jl$ and $\JR$ approximately constant. The black curves depict the separatrix (calculated via perturbation theory, Appendix \ref{sec:app_librationaction}) at the time of trapping (dashed) and at the final time (solid). 

The conservation of the libration action $\Jl$ has an interesting consequence: Since the phase space volume of the resonance grows as the bar slows down, which is directly visible in Fig.~\ref{fig:single_slowAA} as the increase in volume occupied by the separatrix, the gap between the separatrix and the trapped orbits grows. Due to Liouville's theorem, this interspace must be filled with newly trapped stars, and so the resonance builds up layers of trapped stars like a tree grows rings at its bark.

\begin{figure}
  \begin{center}
    \includegraphics[width=8.5cm]{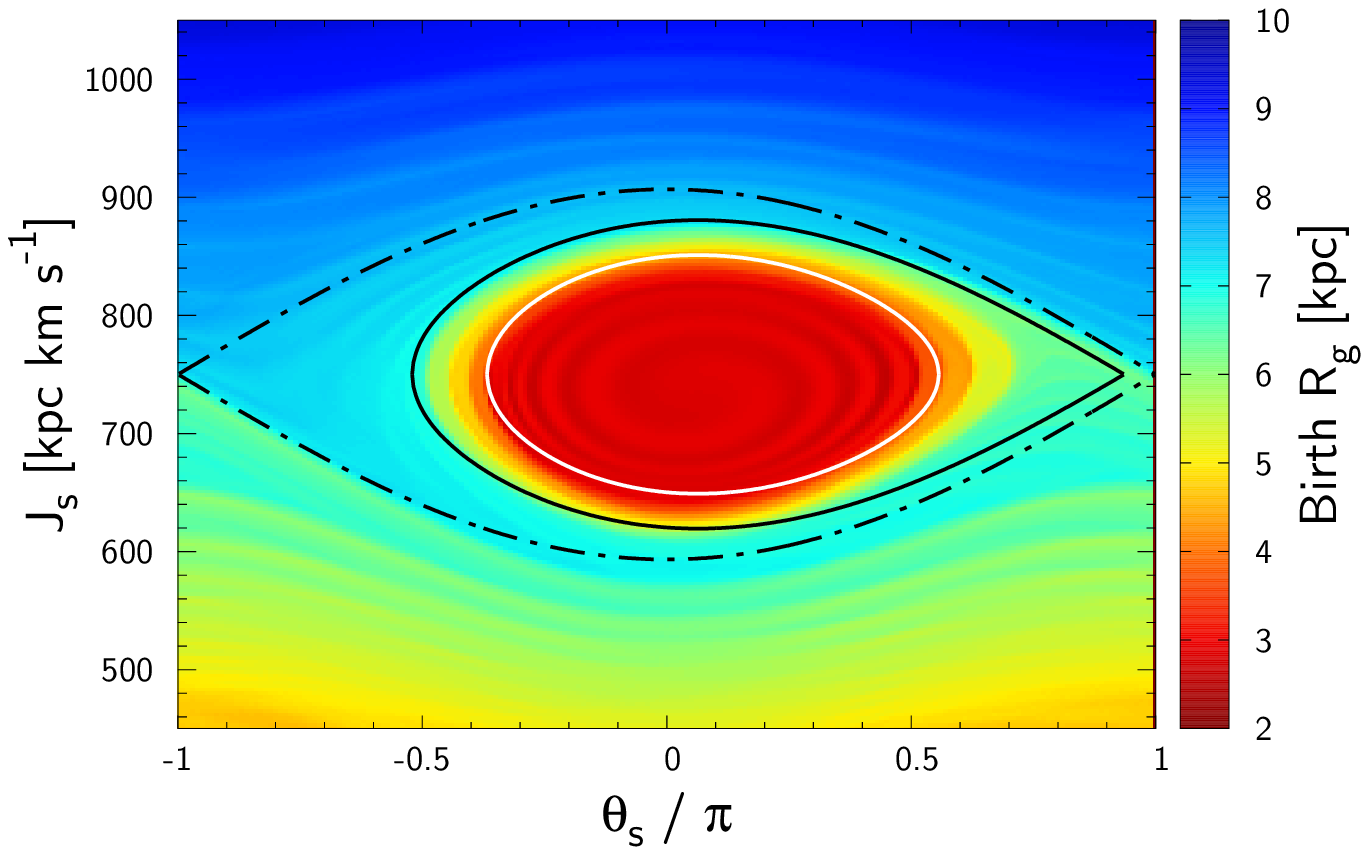}
    \includegraphics[width=8.5cm]{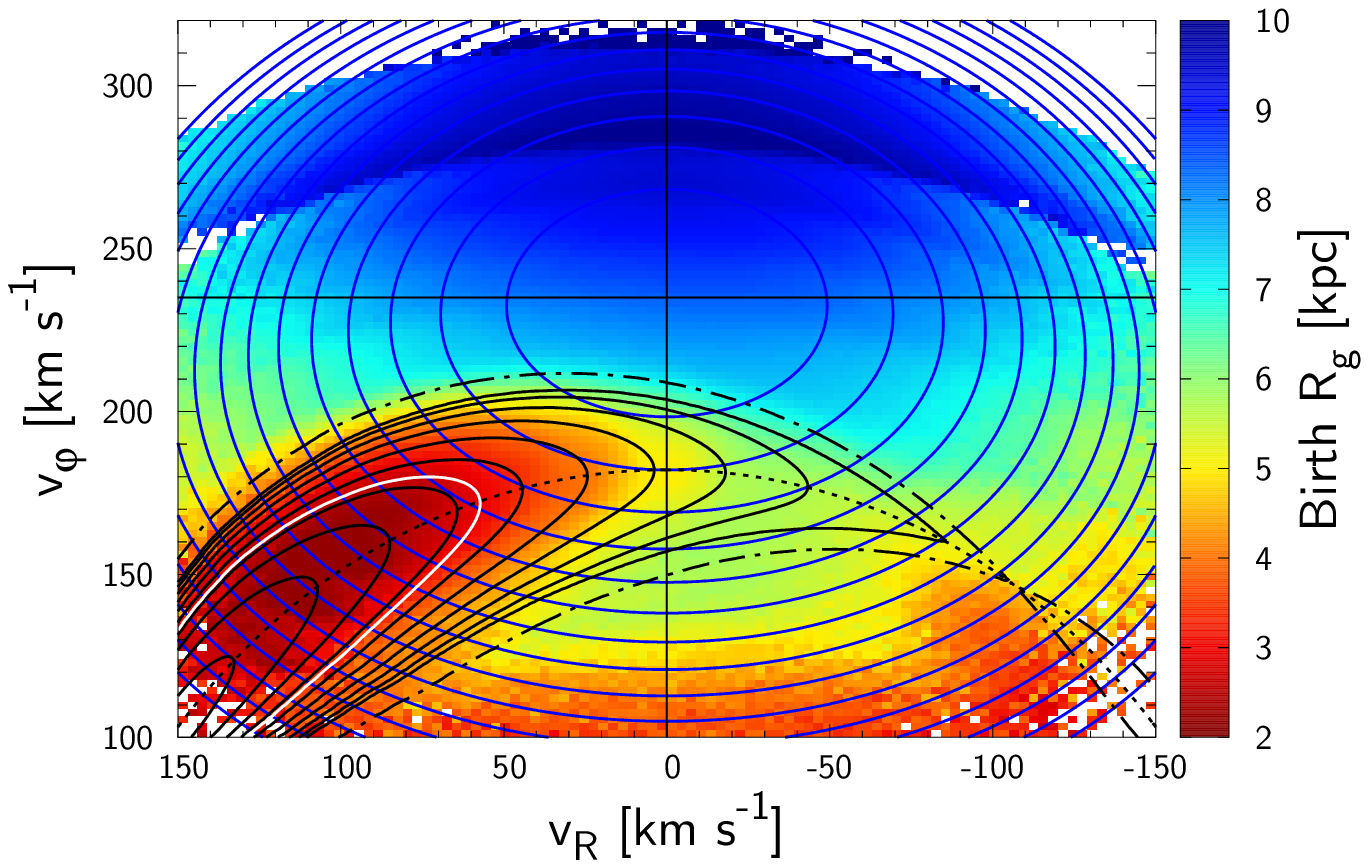}
    \caption{Test-particle simulation of the Galactic disc perturbed by a decelerating bar. Upper panel: Mean birth guiding radius in the slow angle-action plane at $\JR = 10 \kpckms$. The white line marks the contour of constant libration action with a value equivalent to that at the separatrix of the initial bar, while the black solid and dot-dashed curves mark the current separatrix of the moving and resting resonance, respectively. Lower panel: Mean birth guiding radius in local velocity space. The blue and black curves are contours of constant $\JR$ and $\Jln$.}
    \label{fig:simulation_slowAA}
  \end{center}
\end{figure}

As demonstrated with our test particle simulation in the upper panel of Fig.~\ref{fig:simulation_slowAA}, this separates the final resonance into two regimes: the core and the growth region. The initial core, marked by the white curve mapping the original separatrix onto the current volume, contains a relatively homogeneous population trapped initially from further inside the disc at the formation of the bar. Measuring the core can inform us about the initial size and location of the resonance. Between the core and the current separatrix (solid black) lies the growth region where the birth radius monotonically increases towards the separatrix as the fresh layers accreted at later times and thus at larger angular momentum.

Note that the initial formation of the bar (and also the subsequent slow-down) introduce a slight phase-space spiral inside the resonance. Since the period at the separatrix is infinite, the number of wrappings reflects the number of libration periods at the core of the resonance since bar formation. The shape of the phase-space spiral could be used to constrain the change in libration period, which is determined by the shape of the effective potential of the resonance, and ultimately give an estimate on the age of the bar. However, at the current level of \textit{Gaia} data this pattern is not yet detectable.

The black dot-dashed curve shows the separatrix calculated at fixed pattern speed, while the black solid curve takes into account the contraction of the separatrix due to the deceleration (Appendix \ref{sec:app_librationaction}). Only stars within the latter are bound to the resonance, while stars in between are either in transit between the circulating zones or are becoming caught by the resonance.

To test this on observational data, we need to overcome two observational challenges: (i) the Sun is far from the Lagrange points and the available sample from \textit{Gaia} only touches the outskirts of the corotation resonance so currently we can only see stars of relatively large $\Jl$ or $\JR$ that travel far enough from the Lagrange point to reach the Solar neighbourhood. The lower panel of Fig.~\ref{fig:simulation_slowAA} shows how resonances can be identified in the velocity plane of radial vs. azimuthal velocity at the Solar neighbourhood, again coloured by the original guiding centre radii of stars. To guide the eye, we superposed contours of constant $\JR$ in blue equally spaced by $30 \kpckms$, and $\Jl$ normalized by the value at the separatrix $(\Jln \equiv \Jl/\Jlsep)$ in black with $0.1$ spacing. The outermost black curve corresponds to the separatrix (as in the upper panel, dot-dashed for a {\it resting} resonance and solid for a {\it moving} resonance). The dotted black arch represents the location where the resonance condition is exactly satisfied. Although the centre of the resonance cannot be observed, we can see the mean birth radius decreasing towards the initial core of the resonance marked by the white curve. (ii) In practice, we cannot measure the stellar position at trapping, so we employ the metallicity gradient of the Galactic disc: the metallicity of stars increases towards the Galactic centre, so metallicity is a proxy for radius at trapping with one caveat: older stars tend to be more metal poor and have more eccentric orbits/larger radial action, so we need to filter this out when investigating the metallicity trends in $\Jl$.

\section{Method of metallicity estimation}
\label{sec:method}

The metallicity of main sequence stars can be inferred from the position across the main sequence in the colour-magnitude diagram. In general, stars in the main sequence with higher metallicity appear redder. The reason of reddening is two-fold: (i) metals enhance the internal opacity of the star which impedes radiative transport and hence forces the star to swell up with a lower surface temperature. (ii) metals have most of their absorption lines in the UV-blue region. We restrict our analysis to stars sufficiently low on the main sequence so that their colours and magnitudes do not vary significantly with stellar age. In this region, the colour-magnitude position directly encodes the metallicity of a star, with some contamination from binaries and extinction (the extinction vector runs almost parallel to the main sequence, limiting the impact of reddening uncertainties on metallicity estimates).

\subsection{Sample selection}
\label{sec:sample_selection}

We use stellar samples in the Solar neighbourhood (distance from Sun $s < 0.3 \kpc$) taken from the \textit{Gaia} DR2 RV catalogue \citep{GaiaDR2GaiaCollaboration,GaiaDR2Cropper2018,GaiaDR2Sartoretti2018,GaiaDR2Katz2019} with parallax offset and distance estimation from \cite{Schoenrich2019distance}. We adopt the Solar Galactocentric radius $R_0 = 8.2 \kpc$ \citep{Gravity2019geometric}, Solar Galactocentric azimuth with respect to the bar major axis $30^\circ$ \citep{wegg2015structure}, Solar distance from the disc plane $\zsun = 0.02 \kpc$ \citep{Joshi07}, and Solar velocity $(\vRsun, \vphisun - \vc, \vzsun) = (-11.1, 12.24, 7.25) \kms$ \citep{SBD}. We apply quality cuts on parallax $p/\sigpar > 10$ and restrict samples to those with Galactic latitude $\gb > 10^\circ$ to minimize the reddening effect by interstellar extinction. As the resonance lines are expected (and measured, see \citealt{friske2019more}) to exert a mild drift with vertical energy, we exclude stars having $E_z > 200 \kmskms$ corresponding to a maximum vertical velocity of $\vz = 20 \kms$ and a maximum vertical excursion from the Galactic plane of $z \sim 0.3 \kpc$. The vertical potential is evaluated using the Milky Way model of \cite{McMillan2017mass}.

\subsection{Colour-magnitude diagram}
\label{sec:cmd}

\begin{figure}
  \begin{center}
    \includegraphics[width=8.5cm]{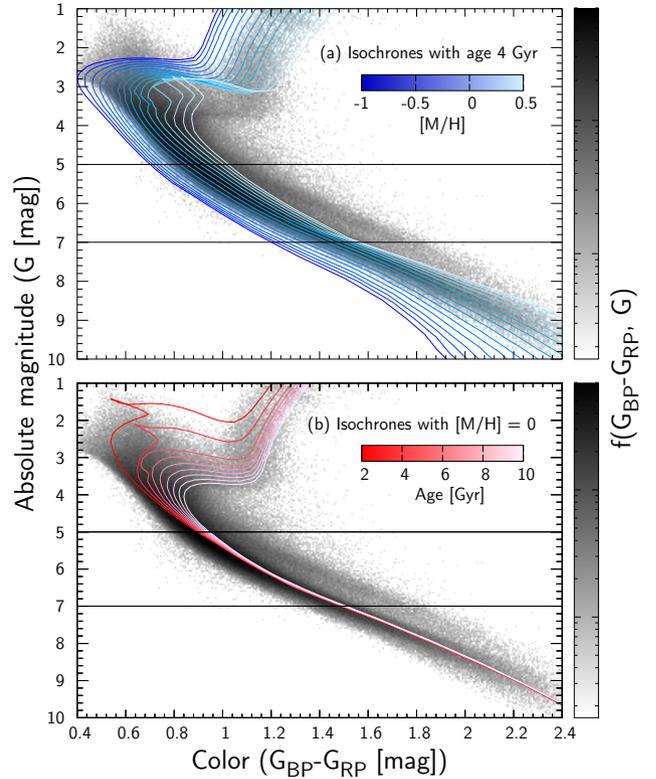}
    \caption{Colour-magnitude diagram of the selected stellar samples in the Solar Neighbourhood. 
    The upper panel overlays isochrones with different metallicities at fixed age $4 \Gyr$, while the lower panel plots isochrones with different age at fixed metallicity $\MH = 0$. The binary sequence is visibly detached above the $\MH = 0.5$ isochrone, where we apply the upper metallicity cut. The two black horizontal lines show our upper/lower limit in magnitude $G$.}
    \label{fig:CMD}
  \end{center}
\end{figure}

Figure \ref{fig:CMD} shows the colour-magnitude diagram of the selected samples superposed by stellar isochrones with (a) fixed age ($4 \Gyr$, blue) and (b) fixed metallicity ($\MH = 0$, red) constructed using PARSEC version 1.2S \citep{Bressan2012PARSEC} (\textit{Gaia} passbands taken from \citealt{Weiler2018Revised}). As discussed at the beginning of this section, an increase in metallicity shifts the isochrones redwards, while age dependence only take an effect near the turn-off region, i.e. on the blue/bright end. To estimate the metallicity of individual stars, we generate isochrones with fixed age ($4 \Gyr$ by default) for metallicities $\MH$ between $-1.0$ and $0.5 \dex$ in $0.05 \dex$ increments, and linearly interpolate them in magnitude $G$. When evaluating the mean metallicity, we cut samples at $G < 5$ near the main-sequence turnoff point. We also apply an upper limit in magnitude ($G < 7$) since the selection function in distance becomes more skewed with increasing magnitude. Finally, we discard samples with metallicity beyond $\MH = 0.5$ since they are most likely to be binary/double stars mistaken for a bright single star.

\subsection{Calibration of photometric metallicity}
\label{sec:calibration}

\begin{figure}
  \begin{center}
    \includegraphics[width=8.5cm]{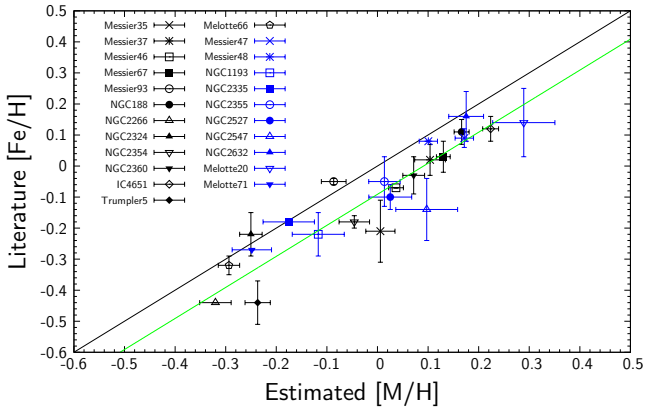}
    \caption{Metallicity of open clusters measured by our method compared with literature values \citep{Netopil2016opencluster}. Data points without vertical error bars are those lacking uncertainty reports in the reference. Open clusters with more than 100 stellar samples are shown in black while those with less than 100 but more than 10 sample stars are shown in blue. The green line, linearly fitted to the data, quantifies the offset $0.091 \pm 0.017 \dex$ of our metallicity estimation due to contamination from binary/double stars.}
    \label{fig:opencluster_MHvsFeH}
  \end{center}
\end{figure}

Due to the contamination by binary/double stars, the metallicity estimated from photometry is generally biased towards high metallicity. To quantify this bias, and to validate our method, we applied our technique to open clusters with metallicities known from spectroscopic measurement. Fig.~\ref{fig:opencluster_MHvsFeH} compares our estimated metallicity $\MH$ with the literature values of $\FeH$ taken from \cite{Netopil2016opencluster}. We select \textit{Gaia} samples within $20 \marcs$ from the cluster core and apply narrow cuts in proper motions around the peak ($\Delta \mu \sim 1 \marcsyr$) to extract members of the clusters. The black data show metallicity inferred from stellar cluster with more than 100 sample stars, while the blue data show those with less than 100 but more than 10 samples. The metallicity estimated from \textit{Gaia} photometry agrees well with that from spectroscopic surveys over a wide range of metallicity up to a constant offset\footnote{Some of the bias may originate from variations in $\alpha$ enhancement \citep{Casagrande2011New}.} ($0.091 \pm 0.017 \dex$) shown in green line (linear fit). The result validates our method and demonstrates that we could compare our results quantitatively with spectroscopic metallicity $\FeH$ by subtracting the constant offset. We caution however that the literature values contain uncertainties beyond the shown error bars as evidenced by the scatter between different catalogues: e.g. Messier 67 has a metallicity of $0.03 \pm 0.05 \dex$ according to \cite{Netopil2016opencluster} but \cite{Carrera2019OpenClusters} reports $0.07 \pm 0.03 \dex$ while \cite{Leaman2012Insights} reports $-0.19 \pm 0.042 \dex$.

\section{Results}
\label{sec:results}

\subsection{Mean metallicity map in local velocity/action space}
\label{sec:local_v_J_space}

\begin{figure*}
  \begin{center}
    \setlength\columnsep{0pt}
    \begin{multicols}{2}
      \includegraphics[width=8.8cm]{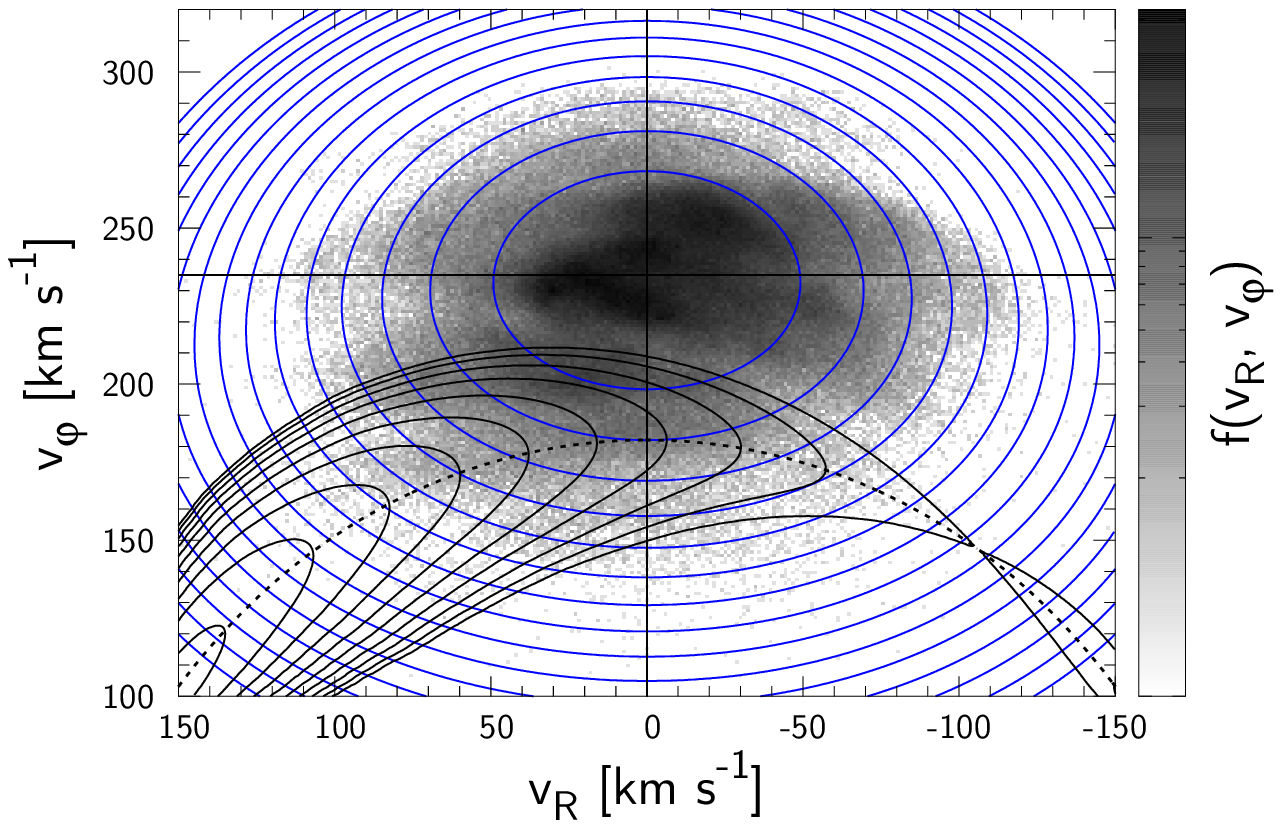}
      \newpage
      \includegraphics[width=8.8cm]{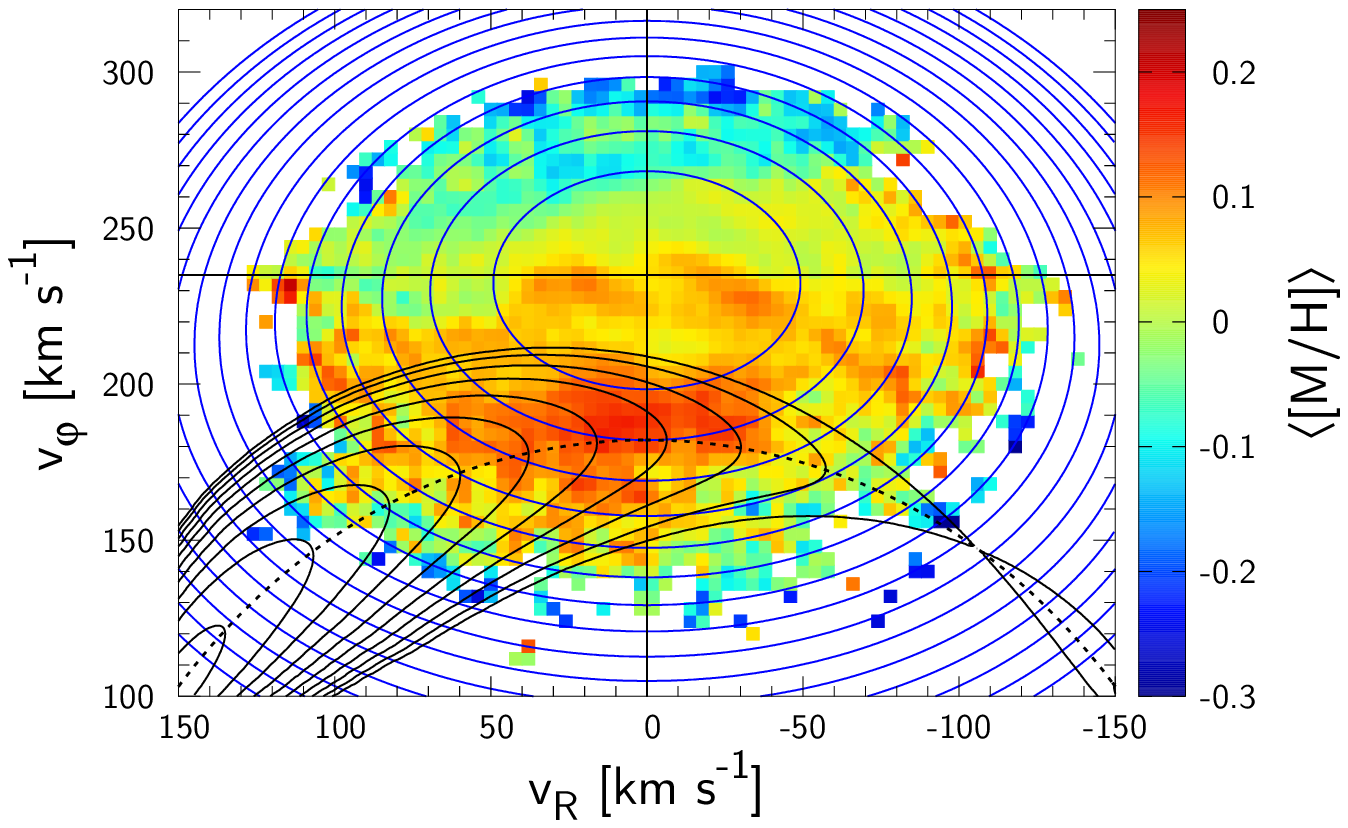}
    \end{multicols}
    \vspace{-0.4cm}
    \caption{Density (left panel) and mean metallicity (right panel) of Solar neighbourhood stars in local velocity plane. Superposed are the uniformly spaced contours of constant $\JR$ in blue and $\Jln\equiv\Jl/\Jlsep$ in black. The dotted black curve marks the location where the resonance condition is exactly satisfied. The mean metallicity is calculated by fitting stellar isochrones to samples in each velocity cell of width $4 \kms$. The Hercules moving group is relatively metal rich indicating an origin at small radii. As we go around the blue contours of constant $\JR$ (in particular the three innermost ellipses), the metallicity rises as we cross the black curves towards small $\Jln$ in agreement with expectation from a decelerating bar model. Note that the colour palette is chosen so that metal-rich/poor stars appear with similar colour as small/large birth $\Rg$ in Fig.~\ref{fig:simulation_slowAA}.}
    \label{fig:v_kin_MH}
  \end{center}
\end{figure*}

\begin{figure*}
  \begin{center}
    \setlength\columnsep{0pt}
    \begin{multicols}{2}
      \includegraphics[width=8.8cm]{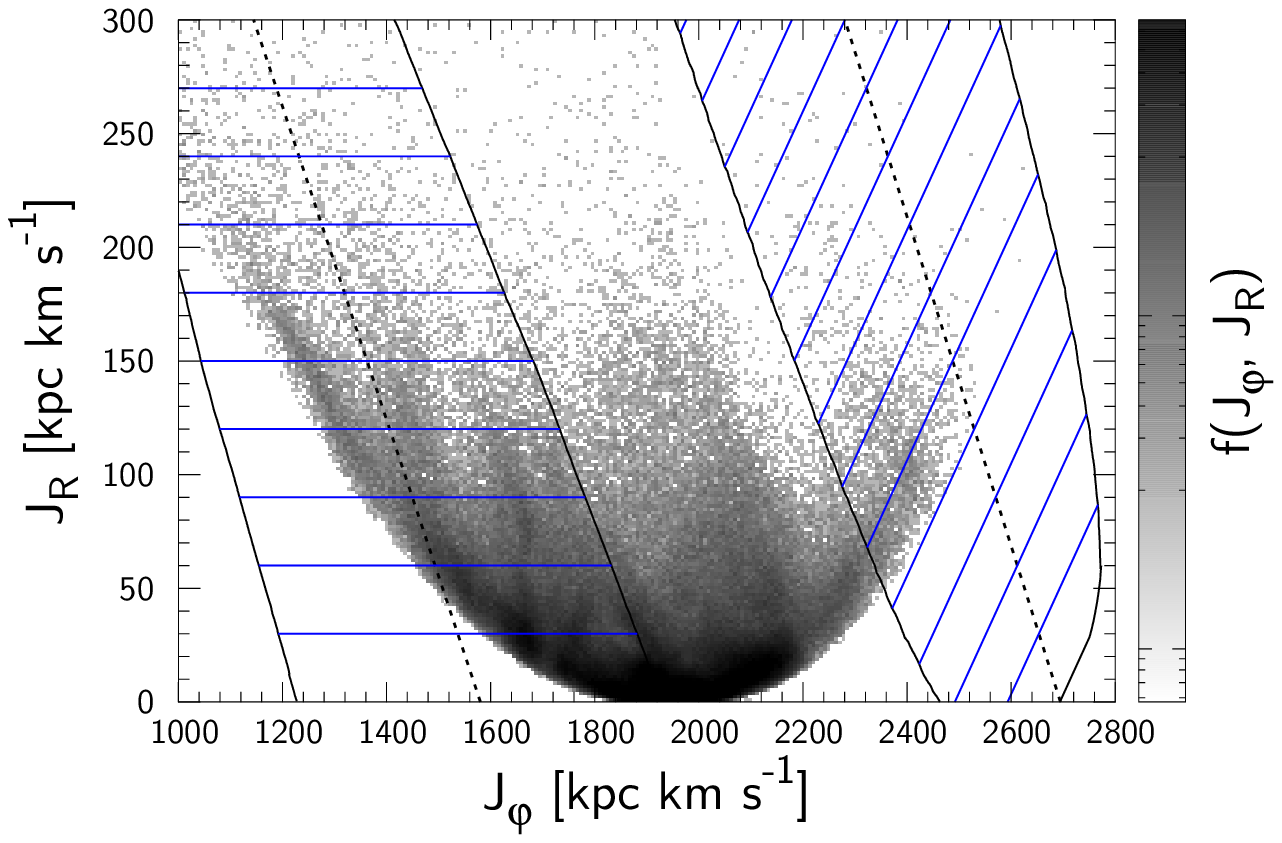}
      \newpage
      \includegraphics[width=8.8cm]{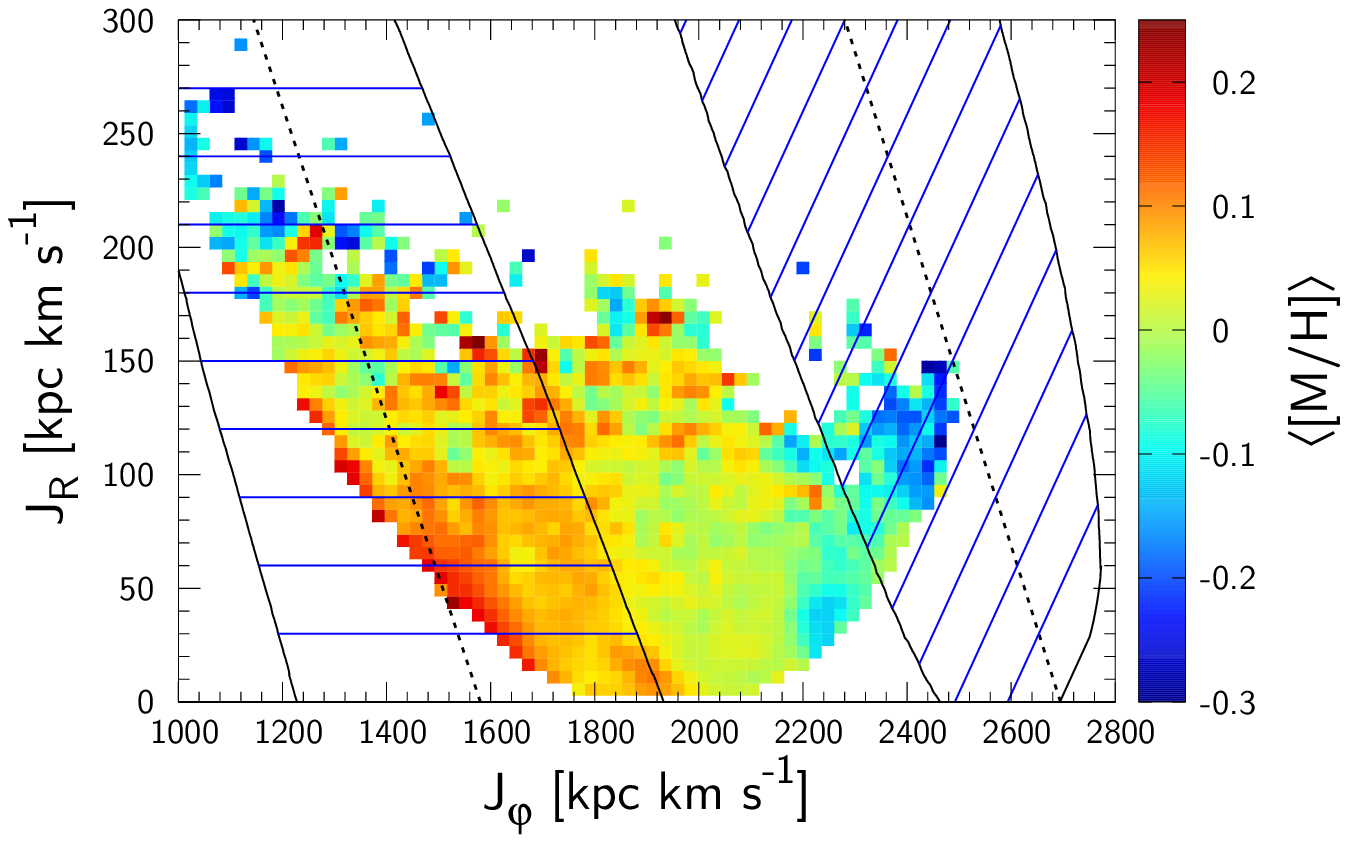}
    \end{multicols}
    \vspace{-0.4cm}
    \caption{Density (left panel) and mean metallicity (right panel) of local stars in action space. The metal poor population appears at high $\Jphi$ (stars visiting the sun from the outer disc) and at high $\JR$ (old stars). The dotted black lines are the CR (left) and the OLR (right) and the solid black lines mark the maximum excursion achieved by trapped orbits librating along contours of constant $\Jf$ (blue lines).}
    \label{fig:J_kin_MH}
  \end{center}
\end{figure*}

The left-hand panel of Fig.~\ref{fig:v_kin_MH} shows the density of stars in the local velocity plane. As in Fig.~\ref{fig:simulation_slowAA}, we overlay contours of constant $\JR$ (blue) and $\Jln$ (black) at $\Omegap = 35 \kmskpc$. The stellar group concentrated around $(\vR,\vphi) \sim (30,190) \kms$ is the `Hercules stream'. Since the studies by \cite{Dehnen1999Pattern,dehnen2000effect}, the origin of the Hercules stream was suspected to be the non-resonant $x_2$ orbits circulating below the outer Lindblad resonance (OLR) of a fast bar ($\Omegap \gtrsim 50 \kmskpc$) \citep[e.g.][]{Antoja2014constraints,fragkoudi2019ridges}. However studies in the past few years have increasingly favoured a slow bar ($\Omegap \lesssim 40 \kmskpc$) in agreement with dynamical models fitted to the kinematics of inner gas \citep{sormani2015gas3} and red clump stars in the bar/bulge region \citep[e.g.][]{Portail2017Dynamical,Clarke2019Milky}. In a slow bar model, the Hercules stream consists of orbits trapped in the corotation resonance (CR) of the bar \citep[e.g.][]{perez2017revisiting,Monari2019signatures,DOnghia2020Trojans,Binney2020Trapped}. This model has the problems that the deformation in the velocity distribution predicted by a constantly rotating bar is not strong enough and less asymmetric in $\vR$ compared to observations. However these problems are resolved by a slowing bar where the CR contracts towards positive $\vR$ and brings stars at high phase-space density from the inner disc \citep{Chiba2020ResonanceSweeping}. There are also models linking the Hercules with the 1:4 resonance of the bar \citep[e.g.][]{Hunt2018OUHR,Asano2020Trimodal} or with transient spiral arms \cite[e.g.][]{Hunt2018Transient}, making the debate indecisive with kinematics only. We will show however that the decelerating slow bar model is singled out by the metallicity trend of Hercules stars. For an extensive comparison between different bar models, see \cite{Trick2021Identifying}.

Figure~\ref{fig:v_kin_MH} right-hand panel colours the local velocity plane in mean metallicity. A similar plot is given by \cite{Antoja2017RAVE} based on spectroscopic metallicity obtained from the RAVE survey and the Geneva-Copenhagen survey. The overall distribution displays the anticipated trends: the decline of metallicity towards larger $\vphi$ (or respectively $\Jphi = \vphi R_0$, where $R_0$ is the Solar Galactocentric radius), reflecting the negative metallicity gradient in Galactocentric radius, and the lower metallicities at larger $\JR$, resulting from the age-metallicity and age-dispersion relationships. However, the most conspicuous feature is the high metallicity clump directly at the position of the Hercules stream. This has already been reported as early as \cite{Grenon1972SMR} from the Geneva photometry and \cite{Grenon1999Kinematics} using the Hipparcos catalogues. Since Hercules is an in-plane stellar stream and is not a dissolved cluster \citep{Bovy2010AHipparcosMovingGroups} nor an accreted population \citep{Kushniruk2020HR1614}, the only natural explanation for its high metallicity is that it originates from the inner Galaxy. This is expected in a decelerating slow bar model since stars trapped in the CR have been dragged from small radii as the bar decelerates \citep{Halle2018Radial,Chiba2020ResonanceSweeping}. In contrast, this observation is unexplainable with a fast bar model where Hercules stars are identified as non-resonant orbits that cannot have been dragged while the resonance swept. The metal-rich nature of Hercules is thus a strong indication that it is composed of resonantly trapped orbits rather than non-trapped orbits. \cite{Antoja2017RAVE} proposed that, in the context of a non-decelerating fast bar model, the high metallicity of Hercules may be explained by the non-resonant $x_2$ orbits which has a slightly smaller mean radius than orbits trapped in the OLR. However, as we show in Appendix \ref{sec:app_metallicity_v_model} using a simple age-metallicity-dispersion relation, a fast bar model is incapable of reproducing the observed high metallicity at the position of the Hercules as the difference in birth radii between trapped and non-trapped orbits is too small (see Fig.~\ref{fig:vRvphiLz0_vRvphiMH}). An analogous argument applies to a non-decelerating slow bar model. Only the decelerating slow bar model can bring the metal rich stars from far inside the disc sufficient to achieve metallicity above $0.2 \dex$ as in the data.

\begin{figure*}
  \begin{center}
    \setlength\columnsep{0pt}
    \begin{multicols}{2}
      \includegraphics[width=8.8cm]{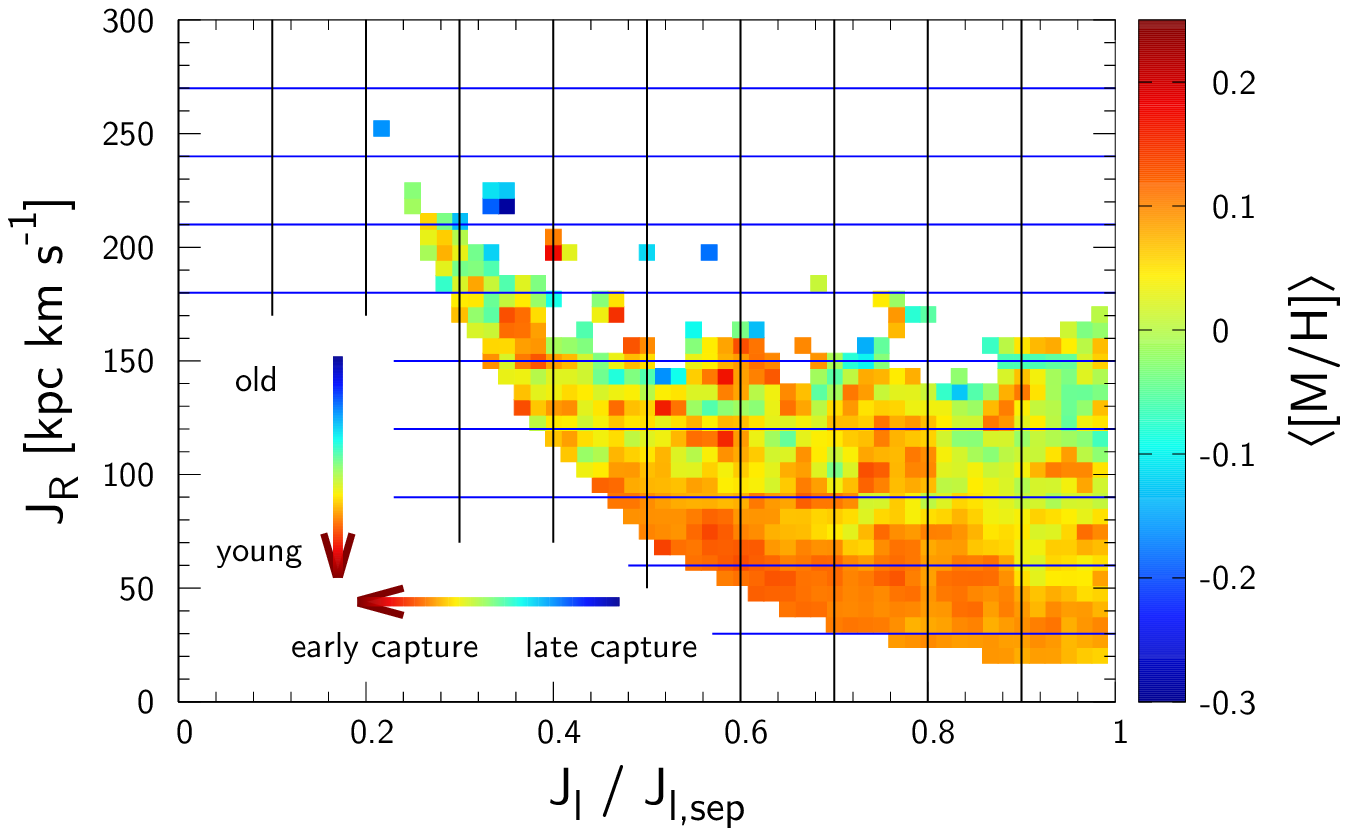}
      \newpage
      \includegraphics[width=8.8cm]{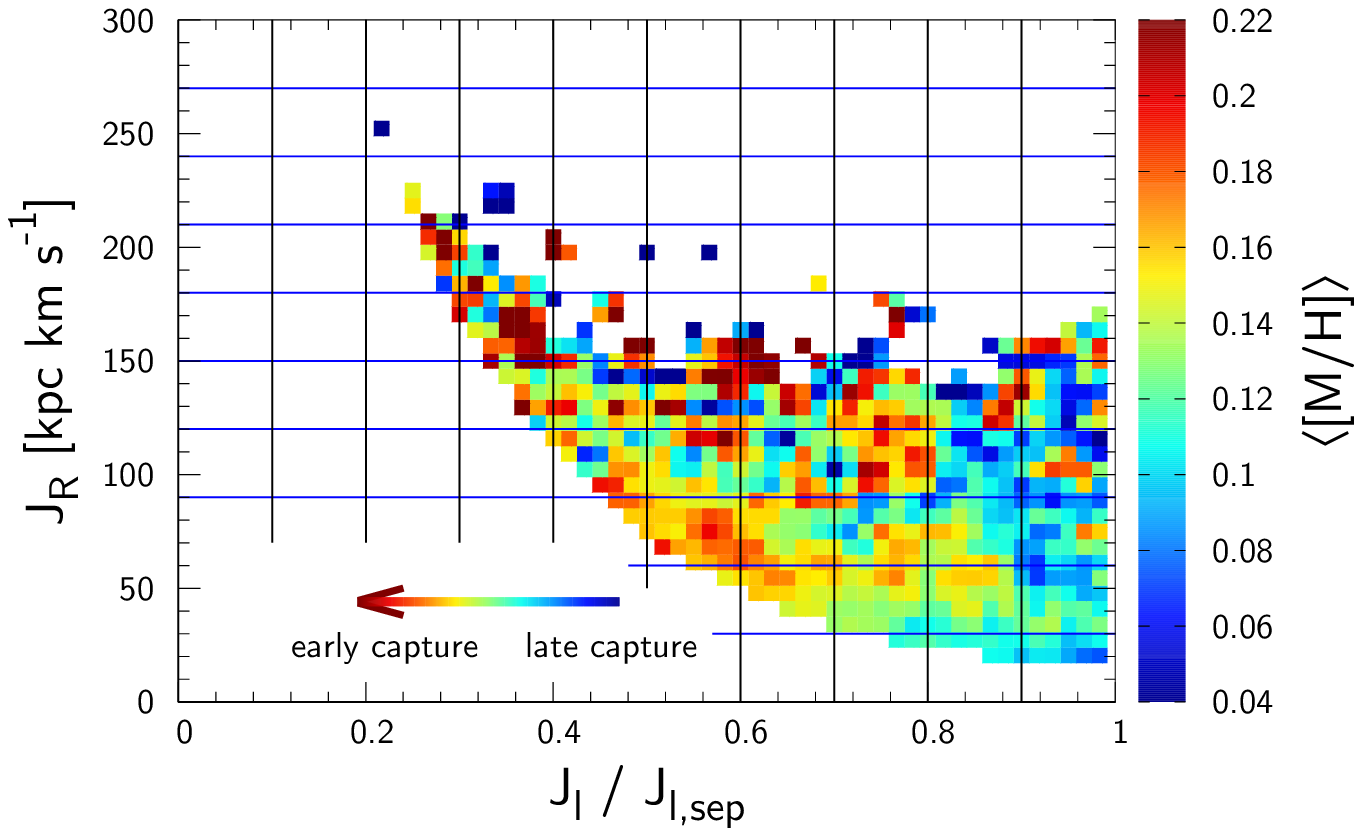}
    \end{multicols}
    \vspace{-0.4cm}
    \caption{Mean metallicity of local stars trapped at the bar's corotation resonance as a function of the resonant actions. Bar pattern speed set to $\Omegap=35 \kmskpc$. Left panel: Metallicity map on resonant actions $\Jln$ and $\JR$. The rectilinear grid corresponds to the contours of constant $\Jln$ and $\JR$ drawn in the local velocity plane (Fig.~\ref{fig:v_kin_MH}). Metallicity is predicted to increase in the directions indicated by the multi-coloured arrows. Right panel: Same plot after removing the metallicity trend in $\JR$ ($\drm \MH / \drm \JR = -0.00078 \dex \kpc^{-1}\,{\rm km}^{-1}\,{\rm s}$) to highlight the metallicity trend in $\Jln$.}
    \label{fig:MH_Jl_JR}
  \end{center}
\end{figure*}

Figure~\ref{fig:J_kin_MH} shows the density and mean metallicity in local action space where the general trend is best observed: the metallicity decreases towards large angular momentum due to the increase in birth radius but also towards large radial action due to the increase in age. There is clearly additional substructure, which can, however, be explained by resonances. Particularly, the metal-rich Hercules stream on the left side of the plot is fitted well with the corotation resonance where the black boundaries mark the the maximum excursion of trapped orbits librating along constant $\Jf=\JR$ (blue lines). At the predicted location of OLR, there is a clear over density (left panel) comprised of metal poor stars (right panel). The distance by which the stars at the OLR can be dragged is limited by the observed radial action; conservation of $\Jf = \JR - \Jphi/2$ implies that the stars acquire a fixed amount of radial action per angular momentum gained, and so the stars observed at e.g. $\JR = 100 \kpckms$ have been dragged in $\Jphi$ by no more than $\Delta\Jphi = 200 \kpckms$ and thus we do not expect high metallicity. In fact, the orientation of resonant dragging in $(\JR,\Jphi)$ space, which is determined by the resonant vector $(\NR,\Nphi)$\footnote{The direction of resonant dragging also depends on the sign of $G\equiv \pd^2 H_0 / \pd \Js^2$ which is identical for resonances at $\NR \geq 0$ \citep{Chiba2020ResonanceSweeping}.}, casts strong limitation on the origin of the Hercules: any outer resonances with $\NR > 0$ will inevitably pump stars up towards larger $\JR$ while dragging them towards the outer disc, so they cannot explain the metal-rich Hercules stars which we observe even at low $\JR$. The only resonance that can carry stars with high metallicity from the inner disc without increasing their eccentricity is the corotation resonance $(\NR = 0)$. Therefore the slow bar is the only model that can explain the metal-rich nature of Hercules using resonantly trapped orbits.

We now map the metallicity onto the resonant actions of the CR to conduct the tree-ring analysis. For this, we have to set parameters for the bar which determine the mapping from $(\vx,\vvel)$ to $(\Jln,\JR)$ for each star within the region of the corotation resonance. The resulting mean metallicity in the $(\Jln,\JR)$ plane is shown in Fig.~\ref{fig:MH_Jl_JR}. The uncertainty of $\Jln$ propagated from the uncertainties in the \textit{Gaia} data is at the percentage level (see Appendix \ref{sec:uncertainty_in_Jl}) so it would not qualitatively affect the signal. The contours of constant $\Jln$ and $\JR$ now form a rectilinear grid. Only stars considered to be inside the resonance enter this plot, so the $x$-axis ranges between 0 (the resonance centre) and 1 (the separatrix). The parabola-like boundary on the left represents the minimum $\JR$ required for trapped orbits to reach the Solar neighbourhood; since orbits with smaller $\Jln$ are confined closer to the Lagrange point, a larger minimum $\JR$ is required to visit us. In the velocity plane, this boundary corresponds to points where the contours of $\Jln$ and $\JR$ are tangential to each other. The mean metallicity increases towards small $\Jln$ (stars captured early at the inner disc) and small $\JR$ (young stars), as indicated by the multi-coloured arrows. To further clarify the metallicity trend in $\Jln$, we show in the right panel the metallicity after subtracting the gradient in $\JR$ obtained by fitting Fig.~\ref{fig:J_kin_MH} right panel with a plane which yields $\drm \MH / \drm \JR = -0.00078 \dex \kpc^{-1}\,{\rm km}^{-1}\,{\rm s}$. We clearly see a monotonic increase towards the resonance centre as predicted for a growing/sweeping resonance, thus implying the slow-down of the bar.

To get a quantitative grip on the observed $\MH$ feature, we project the 2D distribution onto 1D statistics in $\Jl$. Since the sample distribution over $(\Jl,\JR)$ is non-uniform and the $\MH$ depends on $\JR$, a naive averaging over $\JR$ would cause a fatal bias. Therefore, we instead calculate the gradient of $\MH$ with respect to $\Jl$ at each fixed $\JR$ and then average the gradient over $\JR$:
\begin{align}
  \frac{\drm \meanMH}{\drm \Jln}
  &= \frac{\sum_i^{N_{\JR}} w_i \left(\meanMH_i^+ - \meanMH_i^-\right)}{\Delta \Jln \sum_i^{N_{\JR}} w_i}, ~ w_i = \frac{n_i^+ n_i^-}{n_i^+ + n_i^-},
  \label{eq:dmeanMHdJln_weigted}
\end{align}
where the superscript ($\pm$) denotes quantities associated with stars in the bins $\Jln \in [\Jln, \Jln + \Delta \Jln]$ and $\Jln \in [\Jln - \Delta \Jln, \Jln]$. $n_i^{\pm}$ and $\meanMH_i^{\pm}$ are the number of stars and the mean metallicity in the ith $\JR$ bin, respectively. The weights $w_i$ are necessary since each $\JR$ bin contains a different number of stars. Once the metallicity gradient with respect to $\Jln$ is obtained, we reconstruct the metallicity as a function solely of $\Jln$ by integrating the gradient starting from the separatrix. We accumulate the uncertainty of the reconstructed metallicity while taking into account the correlated errors between the metallicity gradients evaluated at adjacent points where they use the same $\Jln$ bin in between. Here we set the bin widths to $\Delta \Jln = 0.05$ and $\Delta \JR = 10 \kpckms$ below which the results do not change significantly.

\begin{figure*}
  \begin{center}
    \setlength\columnsep{0pt}
    \begin{multicols}{2}
      \includegraphics[width=8.8cm]{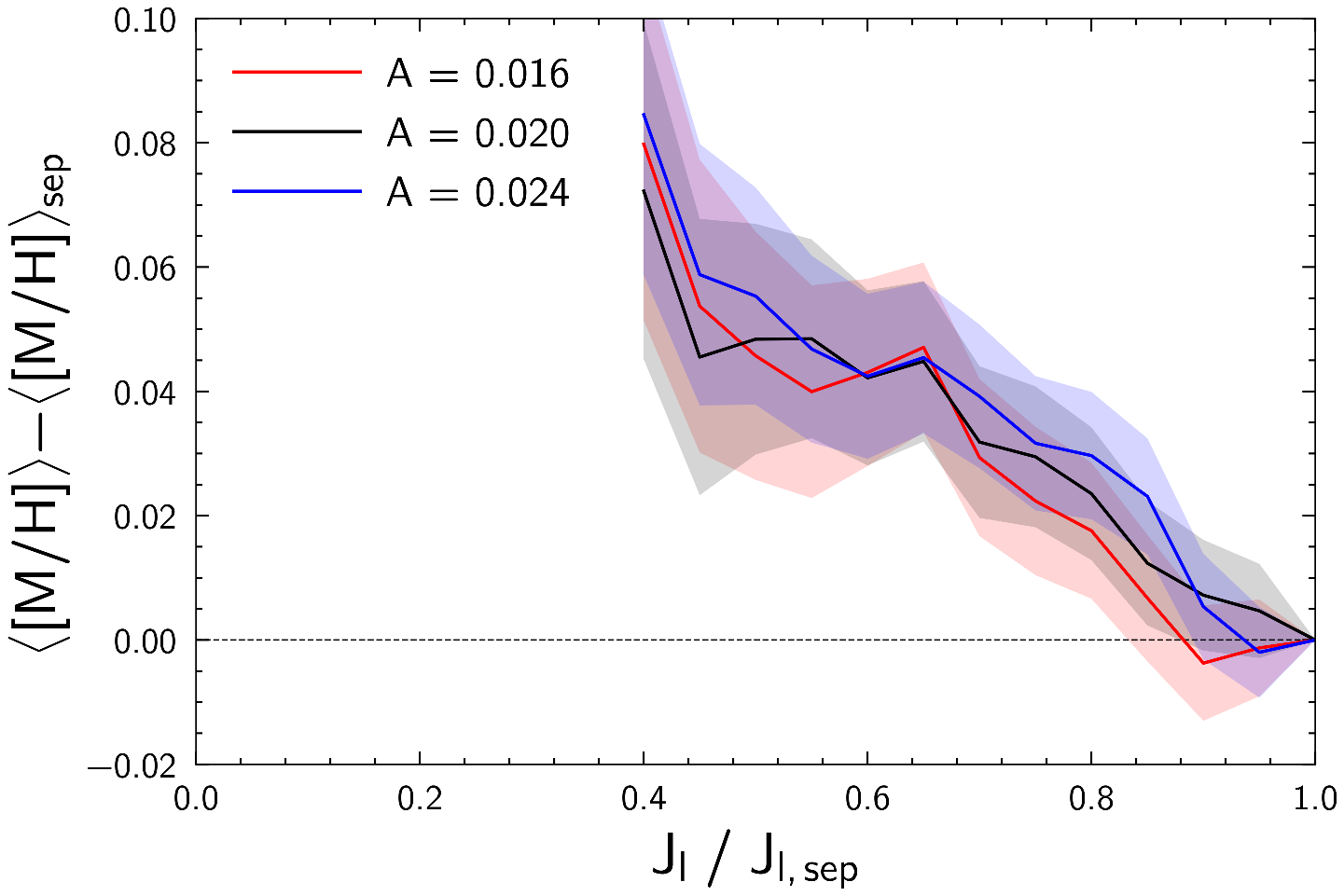}
      \newpage
      \includegraphics[width=8.8cm]{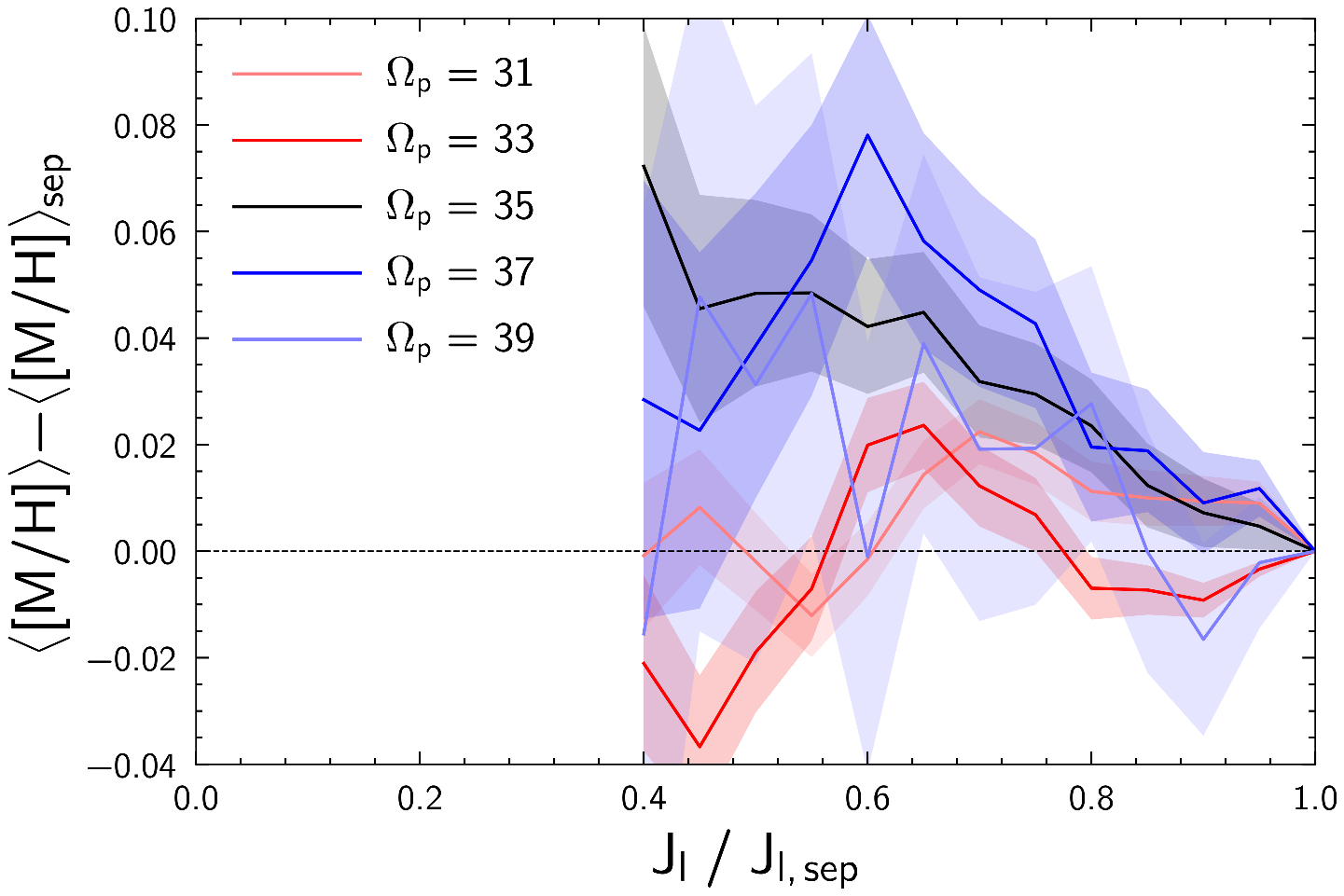}
    \end{multicols}
    \vspace{-0.4cm}
    \caption{Variation of mean metallicity with $\Jln$ derived by integrating the metallicity gradient in $\Jln$. Left panel: Dependence on bar amplitude $A$ at fixed pattern speed $\Omegap = 35 \kmskpc$. At all bar strength, the metallicity decreases monotonically from the centre of the resonance, indicating sequential occupation of the resonance from the core by stars from ever larger radii and thus at lower metallicities. This is the signature of outward migration of the resonance and thus the slow-down of the bar. The coloured bands represent the uncertainties propagated from the 1 s.d. uncertainties of the mean metallicity gradient in $\Jln$. Right panel: Dependence on bar pattern speed $\Omegap$ (units given in $\kmskpc$) at fixed bar strength $A=0.02$. A monotonic trend is only seen with $\Omegap = 35 \kmskpc$ where the location of bar corotation resonance matches the Hercules stream.}
    \label{fig:MH_Jl_A_wp}
  \end{center}
\end{figure*}

The result for this is shown in Fig.~\ref{fig:MH_Jl_A_wp}. The left panel shows the change of mean metallicity with $\Jln$ for a range of bar amplitudes $A$ at fixed pattern speed $\Omegap = 35 \kmskpc$. The signal only weakly depends on the bar strength, i.e. over the whole range of reasonable bar amplitude, we see the same monotonic increase of metallicity towards the resonance centre. As explicitly demonstrated in Appendix \ref{sec:app_metallicity_v_model} (see particularly Fig.~\ref{fig:MH_Jl_mock}) using pseudo-data generated from test-particle simulations, this uptrend in metallicity is only expected when the bar is slowing down. From the total increase in metallicity inside the resonance, we can quantify a lower limit for how much the bar has decelerated (this should turn into a full estimate once \textit{Gaia} can penetrate the core of the resonance). In doing so, we must bear in mind that stars in the disc can be randomly scattered by fluctuations in the gravitational field due to external perturbations (e.g. mergers or satellite interactions) and intrinsic noises (e.g. transient spiral arms or giant molecular clouds), which will tend to weaken the observed metallicity gradient. With this caveat, and given the metallicity gradient of the Galactic disc $-0.05 \dexkpc$ \citep{Luck2018Cepheid}, the maximum increase in the mean metallicity ($\sim 0.08 \dex$) implies that the corotation radius has moved at least $\sim 1.6 \kpc$ outward and thus the pattern speed has declined in excess of $\sim 24\%$ since the formation of the bar.

The right-hand panel of Fig.~\ref{fig:MH_Jl_A_wp} shows the metallicity trend inside the resonance for various bar pattern speeds at fixed bar strength $A = 0.02$. In contrast to the variation in bar amplitude, the pattern speed sensitively affects the metallicity structure: As we decrease the pattern speed, the resonance in the velocity plane shifts towards large $\vphi$ so the metal rich Hercules stars are placed nearer to the lower separatrix of the resonance. Consequently, the metal rich stars become more concentrated at large $\Jl$ and the metallicity wrongly drops towards the inner region of the mis-placed resonance. Similarly, for larger pattern speeds, the upper separatrix of the resonance approaches the metal rich zone and hence the relative increase of metallicity from the separatrix becomes small. A monotonic rise in metallicity towards the core of the resonance is only observed with pattern speed $\Omegap = 35 \kmskpc$ where the CR fits the Hercules stream. This result demonstrates that the slow bar theory is consistent with and strongly favoured by prediction from a decelerating bar.

\subsection{Estimation of bar pattern speed}
\label{sec:pattern_speed}

In the last section we have established the detailed metallicity pattern. We now use its strong $\Omegap$ dependence to measure the bar pattern speed at high precision: The positioning of the resonant actions is only correct with the true pattern speed, and thus the reconstructed metallicity profile will come out of order if we get $\Omegap$ wrong. Since we do not expect the metallicity to undulate against libration action, here we demand a monotonic increase of metallicity towards lower libration action which we quantify by the likelihood of the metallicity to increase at each point in $\Jln$ starting from the separatrix down to $\Jlnmin$. We consider the metallicity change at each point from the previous value
\begin{align}
  &z_i \equiv \meanMH_{i} - \meanMH_{i-1}
  \label{eq:z}
\end{align}
as a random variable distributed normally with its mean $\mzi$ and uncertainty $\szi$ measured. The cumulative distribution function
\begin{align}
  F_i(x) = \frac{1}{2}\left[1 + {\rm erf}\left(\frac{x - \mzi}{\sqrt{2}\szi}\right)\right]
  \label{eq:cumulative}
\end{align}
describes the probability of $z_i$ being smaller than $x$, so the likelihood of $z_i$ being larger than zero is
\begin{align}
  \mathcal{L}_i(A,\Omegap|\mzi,\szi) = 1 - F_i(0) = \frac{1}{2}\left[1 - {\rm erf}\left(\frac{- \mzi}{\sqrt{2}\szi}\right)\right].
  \label{eq:likelihood}
\end{align}
The total likelihood of the metallicity to increase towards the resonance centre is then
\begin{align}
  \mathcal{L}(A,\Omegap|{\bm \mu}_z,{\bm \sigma}_z) = \prod_{i}^{N} \mathcal{L}_i(A,\Omegap|\mzi,\szi),
  \label{eq:likelihood}
\end{align}
where $N \equiv (1 - \Jlnmin) / \Delta \Jln$ is the number of evaluation points in $\Jln$. Since the number of samples drops towards small libration action, the lower limit $\Jlnmin$ is fixed to 0.4 such that, for all bar parameters, each bin in $\Jln$ has more than 30 samples. We have confirmed that the likelihood estimation of \eqref{eq:likelihood} is robust against the choice of $\Delta \Jln$ smaller than 0.1.

\begin{figure}
  \begin{center}
    \includegraphics[width=8cm]{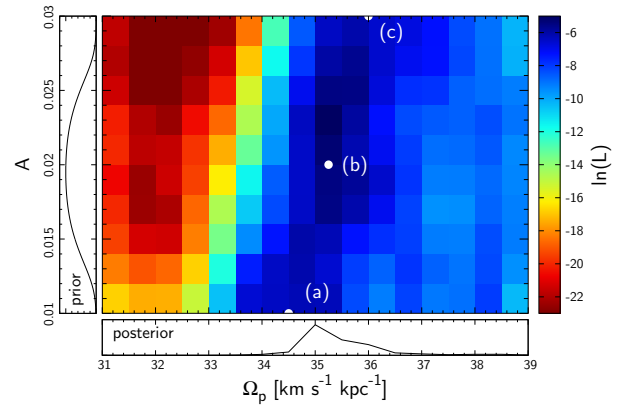}
    \caption{Likelihood (equation \ref{eq:likelihood}) of increase in mean metallicity towards the centre of the bar's corotation resonance plotted over bar pattern speed $\Omegap$ and bar strength $A$. From the prior distribution in $A$ (left panel) inferred from SBM15, we calculate the posterior distribution in $\Omegap$ (bottom panel) which shows a peak at around $\Omegap = 35-36 \kmskpc$.}
    \label{fig:L_A_wp}
  \end{center}
\end{figure}

\begin{figure}
  \begin{center}
      \includegraphics[width=8.5cm]{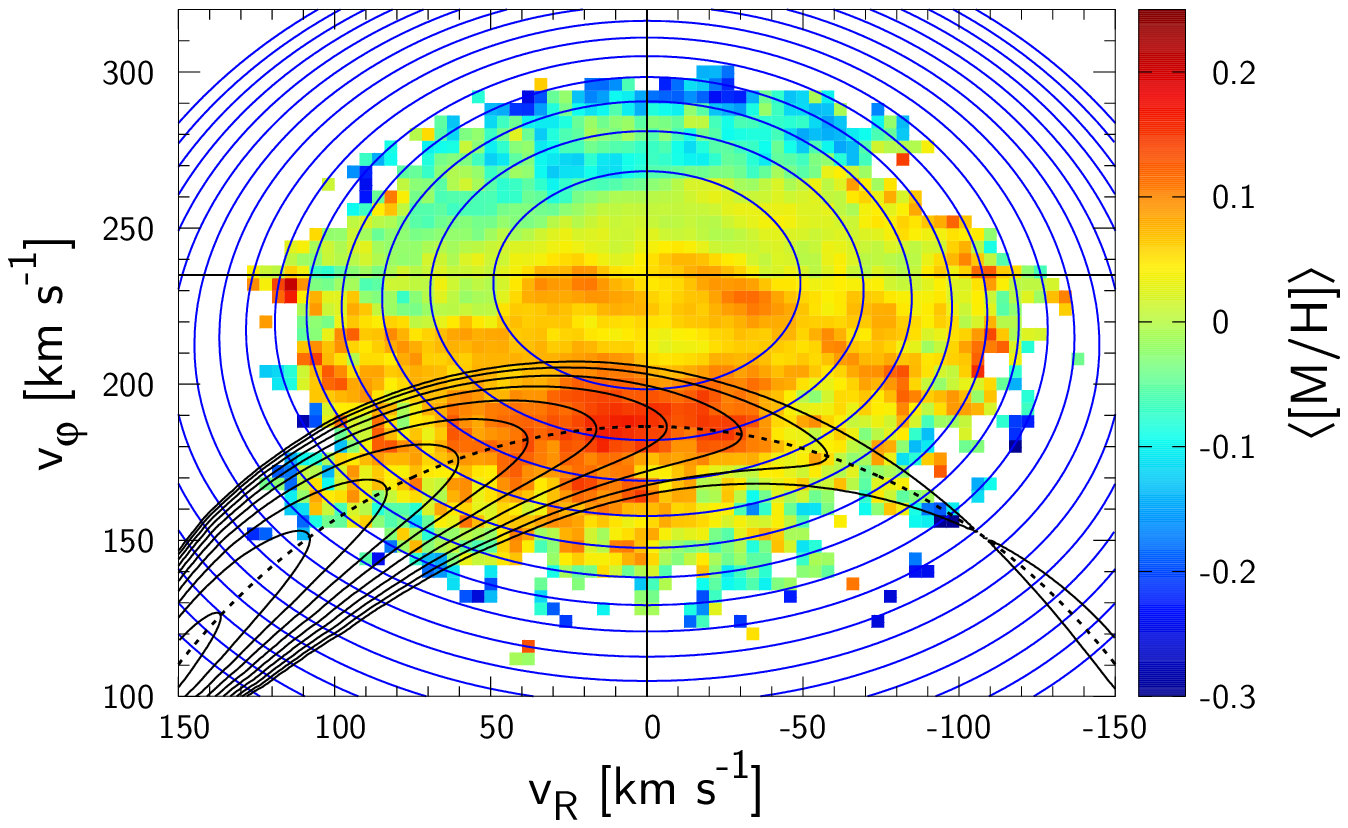}\\
      (a)~$\Omegap = 34.50 \kmskpc, ~ A = 0.01$.\\
      \includegraphics[width=8.5cm]{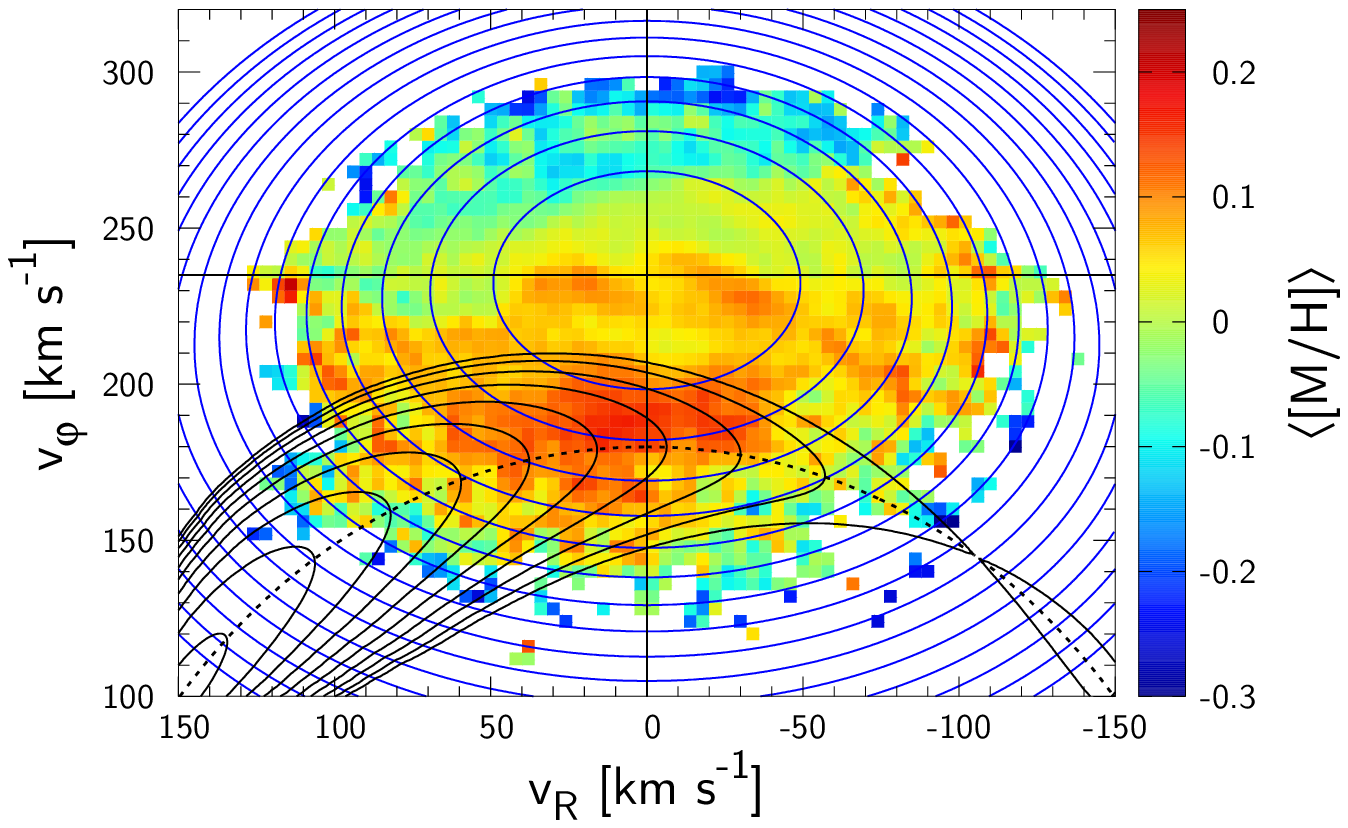}\\
      (b)~$\Omegap = 35.25 \kmskpc, ~ A = 0.02$.\\
      \includegraphics[width=8.5cm]{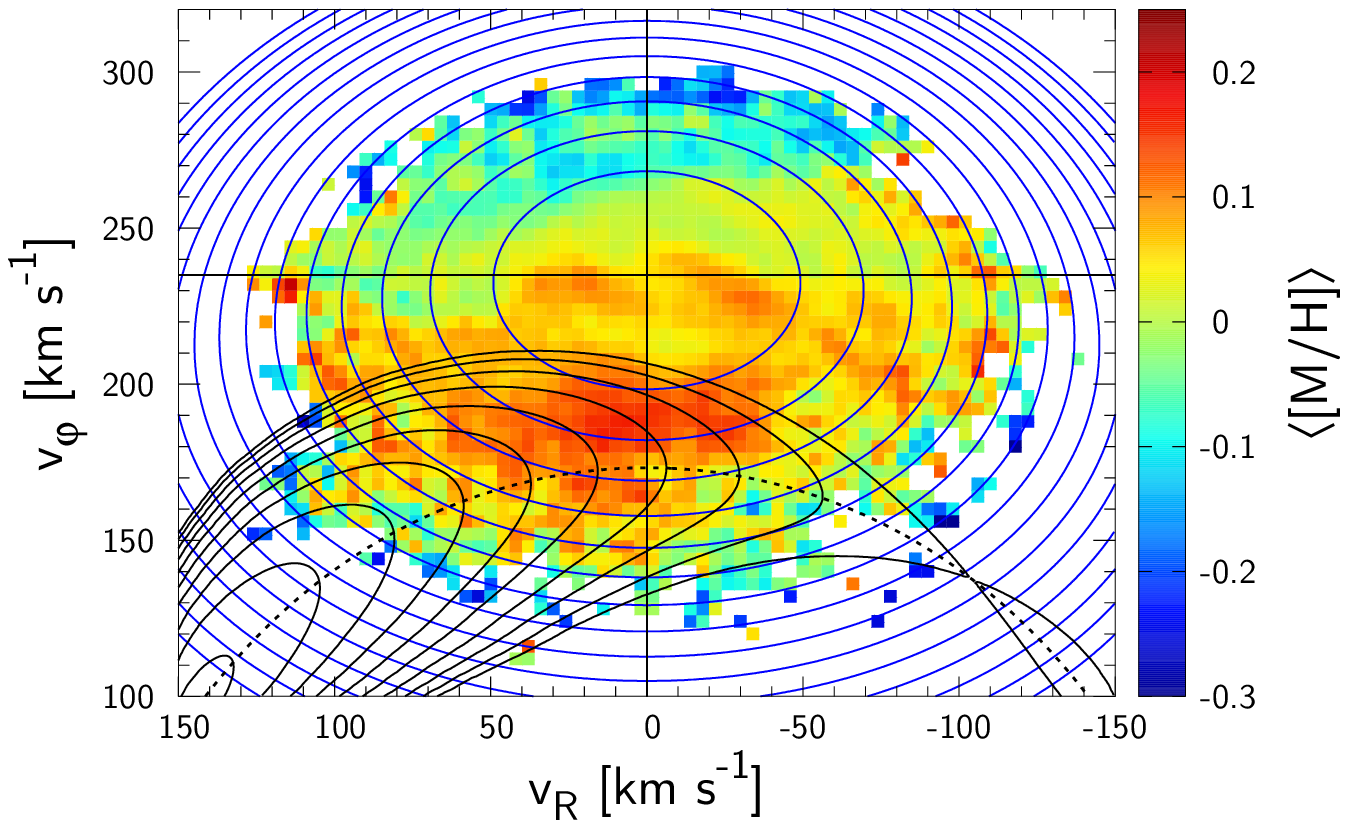}\\
      (c)~$\Omegap = 36.00 \kmskpc, ~ A = 0.03$.
    \caption{Mean metallicity in local velocity space superposed by contours of constant $\JR$ (blue) and $\Jln\equiv\Jl/\Jlsep$ (black) for orbits trapped in the corotation resonance of the bar. All three figures (a)-(c) assume bar parameters (marked on Fig.~\ref{fig:L_A_wp} in white) which yield monotonic increase of metallicity towards small $\Jln$ demanded from the slow-down of the bar.}
    \label{fig:MH_v_Awp}
  \end{center}
\end{figure}

Figure~\ref{fig:L_A_wp} shows the log-likelihood function of the monotonic increase. We observe an inclined peak which means that the faster the rotation of the bar (and thus the lower the location of the resonance in the velocity plane), the more strength (larger resonance size) is required for the metallicity to increase towards the core of the resonance. To have an intuitive understanding of why bar parameters along this inclination are favored in our analysis, we show in Fig.~\ref{fig:MH_v_Awp} the configuration of the resonance in local velocity space for three bar parameters along the peak shown as white circles (a)-(c) in Fig.~\ref{fig:L_A_wp}. From top to bottom, both $\Omegap$ and $A$ increase, i.e. the resonance shifts down but also inflates. Under such changes, the position of the upper separatrix of the CR is kept fixed just above the metal rich zone. As a consequence, for all three figures, the metallicity along constant $\JR$ approximately peaks at points where $\Jln$ is the smallest, i.e. points where contours of $\JR$ and $\Jln$ are tangent to one another, hence resulting in an overall monotonic increase of metallicity towards small $\Jln$. The results simply suggest that bar models with the upper boundary of the CR placed just above the Hercules stream is favoured, in agreement with prediction from kinematics.

From Fig.~\ref{fig:L_A_wp}, we may constrain the bar pattern speed given the priors for the bar amplitude. As in \cite{Chiba2020ResonanceSweeping}, we infer the priors from the study of SBM15. The hydrodynamic simulation by SBM15 suggests that the observed longitude–velocity diagrams of CO and $\rm H_I$ are well reproduced with bar strengths in the range $A_{\rm s} \in [0.4,0.8]$ in their notation which translates to $A \in [0.013,0.026]$ in our model if we fit our analytical bar model to their bar potential beyond half of the corotation radius where the local kinematics are affected by the bar. We assume a normal prior distribution in $A$ with mean $\mu_A = 0.0195$, standard deviation $\sigma_A = 0.0065$, and a smooth cutoff given at $\mu_A \pm \sigma_A$;
\begin{align}
  &P(A) \propto \frac{\exp{\left(- x^2 / 2 \right)}}{\frac{1}{2}\left(\exp{|x^k|} + 1\right)}, ~~~ x \equiv \frac{A-\mu_A}{\sigma_A}
  \label{eq:prior_A}
\end{align}
where $k$ is the cutoff rate set to $k = 4$ as default. $P(A)$ is shown in the left panel of Fig.~\ref{fig:L_A_wp}. The posterior distribution in $\Omegap$ is calculated by integrating $P(A)$ and $\mathcal{L}(A,\Omegap)$ over $A$;
\begin{align}
  P(\Omegap) = \int_{A_{\rm min}}^{A_{\rm max}} dA ~ P(A) \mathcal{L}(A,\Omegap).
  \label{eq:posterior_wp}
\end{align}
Figure~\ref{fig:L_A_wp} bottom panel shows $P(\Omegap)$ which takes mean $\langle \Omegap \rangle = 35.5 \kmskpc$, median $\tilde{\Omega}_{\rm p} = 35.0 \kmskpc$, and standard deviation $\sigma_{\Omegap} = 0.8 \kmskpc$. This is in good agreement with \cite{Binney2020Trapped} who derived $\Omegap = 36 \pm 1 {\Gyr}^{-1} = 35.2 \pm 1.0 \kmskpc$ by applying Jean's theorem to trapped orbits visiting the Sun. Since both studies reached the same conclusion using independent statistics, $\Omegap = 35-36 \kmskpc$ is reliably the optimal pattern speed for the Hercules stream to be composed of orbits trapped in the bar's corotation resonance. 

The pattern speed estimated in this work is slightly lower than recent estimations from stellar kinematics in the bar: both \cite{Sanders2019pattern} and \cite{bovy2019life} estimated $\Omegap = 41 \pm 3\kmskpc$ using the continuity equation. With these intermediate pattern speeds, the upper separatrix of the CR cuts or passes under the metal rich population in the local velocity plane and thus an alternative explanation must be given to the metal rich population outside the bar's corotation resonance. \cite{Portail2017Dynamical} derived $\Omegap = 39 \pm 3.5\kmskpc$ by fitting their dynamical models of the bar region to the density and kinematics of red clump giants using the made-to-measure method, and their models were further used by \cite{Clarke2019Milky} to reproduce the integrated on-sky maps of the longitudinal proper motion which was best achieved at $\Omegap = 37.5 \kmskpc$ in close agreement with our estimation. We conjecture that the small discrepancy is partly due to the uncertainty in the bar strength but more dominantly caused by the difference in the underlying axisymmetric potential.

\subsection{Quantifying systematic uncertainties}
\label{sec:quantify_systematic_uncertainties}

In the following, we discuss the systematic errors of our estimation: Including faint stars by extending the upper limit in \textit{Gaia} magnitude from $G = 7$ to $G = 8$ has a negligible effect, unchanging the optimal bar pattern speed within the reported precision. Changing the age of the isochrone (default $4 \Gyr$) has a marginal impact: with age $6 \Gyr$ we obtain $\Omegap = 35.5 \pm 0.9 \kmskpc$, while with age $2 \Gyr$ we have $\Omegap = 35.4 \pm 0.7 \kmskpc$. Throughout our analysis, we have assumed the angle of bar major axis $\phib$ to be $30^\circ$ ahead from the sun \citep{wegg2015structure}. By varying the bar angle $\phib$, we vary the distance from the sun to the centre of the bar's corotation resonance, i.e. the stable Lagrange point. Therefore, varying the bar angle has an effect on local kinematics similar to varying the bar amplitude. With bar angle of $\phib = 35^\circ$, the optimal pattern speeds rises to $\Omegap = 35.8 \pm 0.9 \kmskpc$. Conversely, decreasing the bar angle to $\phib = 25^\circ$ lowers the best pattern speed down to $\Omegap = 35.2 \pm 0.9 \kmskpc$. We have also analysed the data using resonant actions evaluated in a slowing bar with slowing rate $\eta=0.0036$ as constrained by \cite{Chiba2020ResonanceSweeping}. This yields $\Omegap = 35.5 \pm 1.1 \kmskpc$. The mean is unaffected since the deceleration does not change the location of the resonance, but the uncertainty increases since the resonance contracts reducing the effective sample size. Variation in the axisymmetric potential from a flat circular speed to a slightly inclined one $\vc(R) \equiv (R/R_0)^{~\beta} \vc(R_0)$ has the largest impact changing the pattern speed to $\Omegap = 34.5 \pm 1.1 \kmskpc$ with $\beta = 0.1$ and $\Omegap = 36.8 \pm 0.8 \kmskpc$ with $\beta = -0.1$. 

Finally, uncertainties in $R_0$ and $\vc$ cannot be treated separately as our measurement to first order depends on the local angular frequency $\Omega_0 \equiv \vc/R_0$ which is constrained by the proper motion of Sagittarius ${\rm A}^\ast$, $\mu_{\ell,{\rm A}^{\ast}} = (-6.411\pm0.008)\,\mathrm{mas\,yr^{-1}}=(30.391\pm0.038)\,\mathrm{km\,s^{-1}\,kpc^{-1}}$ \citep{Reid2020SagittariusA} through the relation $v_{\varphi,\odot} = \mu_{\ell,{\rm A}^{\ast}} R_0 = \vc + \Vsun$ and thus
\begin{align}
  \Omega_0 =& \mu_{\ell,{\rm A}^{\ast}} - \Vsun / R_0 \\
  =& 28.90 \pm 0.10 \kmskpc \nonumber
\end{align}
where we take $R_0 = 8.18 \pm 0.02 \kpc$ \citep{Gravity2019geometric} and $\Vsun = 12.24 \pm 0.47 \kms$ \citep{SBD}. There is a systematic uncertainty from wobbles of the black hole against the Galactic centre \citep{Batcheldor2010Displaced} and wobbles of the nuclear region against the large scale disc. These uncertainties amount to a few $\kms$ peculiar motion of Sagittarius ${\rm A}^\ast$ or vice versa a couple per cent in $\mu_{\ell,{\rm A}^{\ast}}$. In this paper, we have assumed $\Omega_0 = \vc / R_0 = 235 \kms / 8.2 \kpc = 28.66 \kmskpc$, so we may be underestimating the pattern speed by a factor of $0.992$ which e.g. shifts our fiducial estimation up to $\Omegap = 35.8 \pm 0.8 \kmskpc$.

\subsection{The relative pattern speed}
\label{sec:relatve_pattern_speed}

Throughout the paper we have referred to the Galactic bar as `slow’ in the sense that its pattern speed is $\Omegap \lesssim 40 \kmskpc$ as opposed to a `fast' bar with $\Omegap \gtrsim 50 \kmskpc$. This slow/fast dichotomy based on the absolute pattern speed is not to be confused with the slow/fast classification based on the dimensionless ratio $\mathcal{R} \equiv \RCR / a_{\rm b}$, where $\RCR$ is the corotation radius and $a_{\rm b}$ is the apparent length of a bar in stellar density. When $1 < \mathcal{R} < 1.4$, the bar is classified as `fast' \citep{athanassoula1992existence,debattista2000constraints,Athanassoula2014Barslowdown}. Most barred galaxies are found to possess a fast bar \citep{Aguerri2015FastBars}. Whether our Galactic bar is fast or slow in the latter sense depends on the measurement of the bar length. \cite{wegg2015structure} fitted the observed red clump stars in the bar region with a parametrized density model and derived $a_{\rm b} = 5.0 \pm 0.2 \kpc$. Adopting their upward revision of the bar length together with our estimation $\RCR = 6.2 \pm 0.2$ yields $\mathcal{R} = 1.3 \pm 0.1$ and thus makes our `slow' pattern speed bar model a typical `fast bar' in this classification.

\section{Conclusions}
\label{sec:conclusion}

We have shown that the resonances of a slowing bar develop like the rings on a growing tree: the distance of trapped orbits to the core of the resonance is adiabatically invariant (i.e. the libration action $\Jl$) and indicates the order of trapping. Since the volume of bar resonance is shown to grow while it sweeps outwards through the disc, newly trapped stars sequentially occupy the phase space near the expanding separatrix. Due to the Galaxy's negative radial metallicity gradient, this pattern is directly observable as a monotonic increase of mean stellar metallicity from the surface towards the core of the resonance.

Using photometric metallicities and stellar kinematics from \textit{Gaia} data, we have shown that the Hercules stream in the Solar Neighbourhood carries this signature. The data displays a highly significant and clean metallicity ordering within the Hercules stream, which we can only explain by identifying Hercules with the bar’s corotation resonance.

The metallicity ordering is only preserved with the correct current bar pattern speed $\Omegap$: The mapping from phase space to the libration action depends critically on $\Omegap$, and so the metallicity ordering in $\Jl$ gets lost at even small changes of $\Omegap$. We showed that this tightly constrains the pattern speed to $\Omegap = 35.5 \pm 0.8 \kmskpc$ and thus $\RCR = 6.6 \pm 0.2 \kpc$, providing another key evidence for the slow bar theory. We stress that a fast bar which associates Hercules with the non-resonant orbits below the outer Lindblad resonance is incompatible with the data since there are no viable mechanism that makes non-trapped orbits significantly metal rich. We further stress that any outer resonances with $\NR > 0$ cannot explain the metal-rich nature of Hercules since resonant dragging in angular momentum will be accompanied by an increase in radial action while the high metallicity stars of Hercules is observed even at low $\JR$. The significant metallicity rise also demands a long sweep in radius, which seems not feasible with a short-lived spiral pattern.

The overall increase in metallicity inside the resonance implies that the corotation radius of the bar must have moved more than $1.6 \kpc$ outwards which corresponds to a decrease of pattern speed by at least $24 \%$ since its formation. A more quantitative understanding on the evolutionary history of the Galactic bar can be gained in the future by fitting the full resonance structure with detailed chemo-dynamical models. Owing to the Sun’s position far from the stable Lagrange points, we currently see only the outer region of the resonance. By performing the analysis at a spatial coordinate closer to the Lagrange points, we could probe deeper into the inner region of the resonance, where we may find traces of events that happened in the early epoch of bar formation (e.g. vertical buckling), and also determine the size of the initial core of the resonance which stems from the formation of the bar. This will be possible in the future with extended data covering the full range of resonance and a proper chemo-dynamical model predicting the age-dependent effects, e.g. the flattening of the radial metallicity gradient towards higher ages by inside-out formation \citep[e.g.][]{Spagna2010,Schoenrich2017Understanding}. We identify further caveats pertaining to diffusion processes. We have not yet evaluated how precisely the structure of a bar-driven resonance will be modified by diffusion processes in phase space due to a variety of perturbations: spiral arms, giant molecular clouds, dwarf galaxy impacts, etc. A naive expectation is that this weakens the metallicity gradient along $\Jl$.

Adding to our previous arguments for a slowing Galactic bar purely based on kinematics, this work provides further evidence using photometry. Hence, our works support the existence of a standard dark-matter halo that has taken up angular momentum from the slowing bar. Alternative theories of gravity are disfavored since they cannot explain the missing angular momentum \citep{Ghafourian2020Modified}. Exotic dark matter in the form of degenerate quantum condensates \citep[e.g.][]{Goodman2000Repulsive,Hu2000Fuzzy}, recently favored by virtue of preventing the formation of density cusps, must be tested for their degree of angular momentum exchange with the baryonic bar. Thus, the discovery of the deceleration of the bar provides a new testbed through which any successful dark matter model must pass. The bar slow-down also paves the path to a new class of constraints on the dark halo: The dynamical friction on the bar depends on both the dark halo's density and kinematics, and thus in combination with standard maps of the gravitational potential, gives us access to measuring the dynamical properties of the dark halo (e.g. rotation).

\section*{Acknowledgements}

We thank J. Binney, D. Kawata, M. Cropper, W. Dehnen, and members of the Oxford galactic dynamics group for valuable discussions. R.C. acknowledges financial support from the Takenaka Scholarship Foundation and the Royal Society grant RGF$\backslash$R1$\backslash$180095. R.S. is supported by a Royal Society University Research Fellowship. This work was performed using the Cambridge Service for Data Driven Discovery (CSD3), part of which is operated by the University of Cambridge Research Computing on behalf of the STFC DiRAC HPC Facility (www.dirac.ac.uk). The DiRAC component of CSD3 was funded by BEIS capital funding via STFC capital grants ST/P002307/1 and ST/R002452/1 and STFC operations grant ST/R00689X/1. DiRAC is part of the National e-Infrastructure. This work has made use of data from the European Space Agency (ESA) mission {\it Gaia} (\url{https://www.cosmos.esa.int/gaia}), processed by the {\it Gaia} Data Processing and Analysis Consortium (DPAC,\url{https://www.cosmos.esa.int/web/gaia/dpac/consortium}). Funding for the DPAC has been provided by national institutions, in particular the institutions participating in the {\it Gaia} Multilateral Agreement.


\section*{Data availability}
This study used the data from {\it Gaia} publicly available at \url{https://gea.esac.esa.int/archive}. The distances and parallax offsets for the Gaia sources are taken from \cite{Schoenrich2019distance} and are available at \url{https://zenodo.org/record/2557803}. The isochrones used to estimate stellar metallicity are generated by PARSEC version 1.2S \citep{Bressan2012PARSEC} at \url{http://stev.oapd.inaf.it/cgi-bin/cmd}. The code used to compute the Galactic potential from \cite{McMillan2017mass} is available at \url{https://github.com/PaulMcMillan-Astro/GalPot}. The codes used to perform the test particle simulations, to compute the angle-action coordinates, and to conduct the data analysis are available from the corresponding author upon request.


\bibliographystyle{mnras}
\bibliography{./bar}

\begin{thebibliography}{}
\makeatletter
\relax
\def\mn@urlcharsother{\let\do\@makeother \do\$\do\&\do\#\do\^\do\_\do\%\do\~}
\def\mn@doi{\begingroup\mn@urlcharsother \@ifnextchar [ {\mn@doi@}
  {\mn@doi@[]}}
\def\mn@doi@[#1]#2{\def\@tempa{#1}\ifx\@tempa\@empty \href
  {http://dx.doi.org/#2} {doi:#2}\else \href {http://dx.doi.org/#2} {#1}\fi
  \endgroup}
\def\mn@eprint#1#2{\mn@eprint@#1:#2::\@nil}
\def\mn@eprint@arXiv#1{\href {http://arxiv.org/abs/#1} {{\tt arXiv:#1}}}
\def\mn@eprint@dblp#1{\href {http://dblp.uni-trier.de/rec/bibtex/#1.xml}
  {dblp:#1}}
\def\mn@eprint@#1:#2:#3:#4\@nil{\def\@tempa {#1}\def\@tempb {#2}\def\@tempc
  {#3}\ifx \@tempc \@empty \let \@tempc \@tempb \let \@tempb \@tempa \fi \ifx
  \@tempb \@empty \def\@tempb {arXiv}\fi \@ifundefined
  {mn@eprint@\@tempb}{\@tempb:\@tempc}{\expandafter \expandafter \csname
  mn@eprint@\@tempb\endcsname \expandafter{\@tempc}}}

\bibitem[\protect\citeauthoryear{{Aguerri} et~al.,}{{Aguerri}
  et~al.}{2015}]{Aguerri2015FastBars}
{Aguerri} J.~A.~L.,  et~al., 2015, \mn@doi [\aap]
  {10.1051/0004-6361/201423383}, \href
  {https://ui.adsabs.harvard.edu/abs/2015A&A...576A.102A} {576, A102}

\bibitem[\protect\citeauthoryear{{Antoja} et~al.,}{{Antoja}
  et~al.}{2014}]{Antoja2014constraints}
{Antoja} T.,  et~al., 2014, \mn@doi [\aap] {10.1051/0004-6361/201322623}, \href
  {https://ui.adsabs.harvard.edu/abs/2014A&A...563A..60A} {563, A60}

\bibitem[\protect\citeauthoryear{{Antoja} et~al.,}{{Antoja}
  et~al.}{2017}]{Antoja2017RAVE}
{Antoja} T.,  et~al., 2017, \mn@doi [\aap] {10.1051/0004-6361/201629387}, \href
  {https://ui.adsabs.harvard.edu/abs/2017A&A...601A..59A} {601, A59}

\bibitem[\protect\citeauthoryear{{Asano}, {Fujii}, {Baba}, {B{\'e}dorf},
  {Sellentin}  \& {Portegies Zwart}}{{Asano} et~al.}{2020}]{Asano2020Trimodal}
{Asano} T.,  {Fujii} M.~S.,  {Baba} J.,  {B{\'e}dorf} J.,  {Sellentin} E.,
  {Portegies Zwart} S.,  2020, \mn@doi [\mnras] {10.1093/mnras/staa2849}, \href
  {https://ui.adsabs.harvard.edu/abs/2020MNRAS.499.2416A} {499, 2416}

\bibitem[\protect\citeauthoryear{{Athanassoula}}{{Athanassoula}}{1992}]{athanassoula1992existence}
{Athanassoula} E.,  1992, \mn@doi [\mnras] {10.1093/mnras/259.2.345}, \href
  {https://ui.adsabs.harvard.edu/abs/1992MNRAS.259..345A} {259, 345}

\bibitem[\protect\citeauthoryear{{Athanassoula}}{{Athanassoula}}{2003}]{athanassoula2003determines}
{Athanassoula} E.,  2003, \mn@doi [\mnras] {10.1046/j.1365-8711.2003.06473.x},
  \href {https://ui.adsabs.harvard.edu/abs/2003MNRAS.341.1179A} {341, 1179}

\bibitem[\protect\citeauthoryear{{Athanassoula}}{{Athanassoula}}{2014}]{Athanassoula2014Barslowdown}
{Athanassoula} E.,  2014, \mn@doi [\mnras] {10.1093/mnrasl/slt163}, \href
  {https://ui.adsabs.harvard.edu/abs/2014MNRAS.438L..81A} {438, L81}

\bibitem[\protect\citeauthoryear{{Batcheldor}, {Robinson}, {Axon}, {Perlman}
  \& {Merritt}}{{Batcheldor} et~al.}{2010}]{Batcheldor2010Displaced}
{Batcheldor} D.,  {Robinson} A.,  {Axon} D.~J.,  {Perlman} E.~S.,   {Merritt}
  D.,  2010, \mn@doi [\apjl] {10.1088/2041-8205/717/1/L6}, \href
  {https://ui.adsabs.harvard.edu/abs/2010ApJ...717L...6B} {717, L6}

\bibitem[\protect\citeauthoryear{{Binney}}{{Binney}}{2020a}]{Binney2020Lindblad}
{Binney} J.,  2020a, \mn@doi [\mnras] {10.1093/mnras/staa092}, \href
  {https://ui.adsabs.harvard.edu/abs/2020MNRAS.495..886B} {495, 886}

\bibitem[\protect\citeauthoryear{{Binney}}{{Binney}}{2020b}]{Binney2020Trapped}
{Binney} J.,  2020b, \mn@doi [\mnras] {10.1093/mnras/staa1103}, \href
  {https://ui.adsabs.harvard.edu/abs/2020MNRAS.495..895B} {495, 895}

\bibitem[\protect\citeauthoryear{{Binney} \& {Tremaine}}{{Binney} \&
  {Tremaine}}{2008}]{binney2008galactic}
{Binney} J.,  {Tremaine} S.,  2008, {Galactic Dynamics: Second Edition}.
Princeton University Press

\bibitem[\protect\citeauthoryear{{Bovy} \& {Hogg}}{{Bovy} \&
  {Hogg}}{2010}]{Bovy2010AHipparcosMovingGroups}
{Bovy} J.,  {Hogg} D.~W.,  2010, \mn@doi [\apj] {10.1088/0004-637X/717/2/617},
  \href {https://ui.adsabs.harvard.edu/abs/2010ApJ...717..617B} {717, 617}

\bibitem[\protect\citeauthoryear{{Bovy}, {Leung}, {Hunt}, {Mackereth},
  {Garc{\'\i}a-Hern{\'a}ndez}  \& {Roman-Lopes}}{{Bovy}
  et~al.}{2019}]{bovy2019life}
{Bovy} J.,  {Leung} H.~W.,  {Hunt} J. A.~S.,  {Mackereth} J.~T.,
  {Garc{\'\i}a-Hern{\'a}ndez} D.~A.,   {Roman-Lopes} A.,  2019, \mn@doi
  [\mnras] {10.1093/mnras/stz2891}, \href
  {https://ui.adsabs.harvard.edu/abs/2019MNRAS.490.4740B} {490, 4740}

\bibitem[\protect\citeauthoryear{{Bressan}, {Marigo}, {Girardi}, {Salasnich},
  {Dal Cero}, {Rubele}  \& {Nanni}}{{Bressan} et~al.}{2012}]{Bressan2012PARSEC}
{Bressan} A.,  {Marigo} P.,  {Girardi} L.,  {Salasnich} B.,  {Dal Cero} C.,
  {Rubele} S.,   {Nanni} A.,  2012, \mn@doi [\mnras]
  {10.1111/j.1365-2966.2012.21948.x}, \href
  {https://ui.adsabs.harvard.edu/abs/2012MNRAS.427..127B} {427, 127}

\bibitem[\protect\citeauthoryear{{Carrera} et~al.,}{{Carrera}
  et~al.}{2019}]{Carrera2019OpenClusters}
{Carrera} R.,  et~al., 2019, \mn@doi [\aap] {10.1051/0004-6361/201834546},
  \href {https://ui.adsabs.harvard.edu/abs/2019A&A...623A..80C} {623, A80}

\bibitem[\protect\citeauthoryear{{Casagrande}, {Sch{\"o}nrich}, {Asplund},
  {Cassisi}, {Ram{\'\i}rez}, {Mel{\'e}ndez}, {Bensby}  \&
  {Feltzing}}{{Casagrande} et~al.}{2011}]{Casagrande2011New}
{Casagrande} L.,  {Sch{\"o}nrich} R.,  {Asplund} M.,  {Cassisi} S.,
  {Ram{\'\i}rez} I.,  {Mel{\'e}ndez} J.,  {Bensby} T.,   {Feltzing} S.,  2011,
  \mn@doi [\aap] {10.1051/0004-6361/201016276}, \href
  {https://ui.adsabs.harvard.edu/abs/2011A&A...530A.138C} {530, A138}

\bibitem[\protect\citeauthoryear{{Chiba}, {Friske}  \& {Sch{\"o}nrich}}{{Chiba}
  et~al.}{2020}]{Chiba2020ResonanceSweeping}
{Chiba} R.,  {Friske} J. K.~S.,   {Sch{\"o}nrich} R.,  2020, \mn@doi [\mnras]
  {10.1093/mnras/staa3585}, \href
  {https://ui.adsabs.harvard.edu/abs/2020MNRAS.tmp.3375C} {}

\bibitem[\protect\citeauthoryear{{Clarke}, {Wegg}, {Gerhard}, {Smith}, {Lucas}
  \& {Wylie}}{{Clarke} et~al.}{2019}]{Clarke2019Milky}
{Clarke} J.~P.,  {Wegg} C.,  {Gerhard} O.,  {Smith} L.~C.,  {Lucas} P.~W.,
  {Wylie} S.~M.,  2019, \mn@doi [\mnras] {10.1093/mnras/stz2382}, \href
  {https://ui.adsabs.harvard.edu/abs/2019MNRAS.489.3519C} {489, 3519}

\bibitem[\protect\citeauthoryear{{Collett}, {Dutta}  \& {Evans}}{{Collett}
  et~al.}{1997}]{Collett1997Capture}
{Collett} J.~L.,  {Dutta} S.~N.,   {Evans} N.~W.,  1997, \mn@doi [\mnras]
  {10.1093/mnras/285.1.49}, \href
  {https://ui.adsabs.harvard.edu/abs/1997MNRAS.285...49C} {285, 49}

\bibitem[\protect\citeauthoryear{{Cropper} et~al.,}{{Cropper}
  et~al.}{2018}]{GaiaDR2Cropper2018}
{Cropper} M.,  et~al., 2018, \mn@doi [\aap] {10.1051/0004-6361/201832763},
  \href {https://ui.adsabs.harvard.edu/abs/2018A&A...616A...5C} {616, A5}

\bibitem[\protect\citeauthoryear{{D'Onghia} \& {L. Aguerri}}{{D'Onghia} \& {L.
  Aguerri}}{2020}]{DOnghia2020Trojans}
{D'Onghia} E.,  {L. Aguerri} J.~A.,  2020, \mn@doi [\apj]
  {10.3847/1538-4357/ab6bd6}, \href
  {https://ui.adsabs.harvard.edu/abs/2020ApJ...890..117D} {890, 117}

\bibitem[\protect\citeauthoryear{{Debattista} \& {Sellwood}}{{Debattista} \&
  {Sellwood}}{2000}]{debattista2000constraints}
{Debattista} V.~P.,  {Sellwood} J.~A.,  2000, \mn@doi [\apj] {10.1086/317148},
  \href {https://ui.adsabs.harvard.edu/abs/2000ApJ...543..704D} {543, 704}

\bibitem[\protect\citeauthoryear{{Dehnen}}{{Dehnen}}{1999}]{Dehnen1999Pattern}
{Dehnen} W.,  1999, \mn@doi [\apjl] {10.1086/312299}, \href
  {https://ui.adsabs.harvard.edu/abs/1999ApJ...524L..35D} {524, L35}

\bibitem[\protect\citeauthoryear{{Dehnen}}{{Dehnen}}{2000}]{dehnen2000effect}
{Dehnen} W.,  2000, \mn@doi [\aj] {10.1086/301226}, \href
  {https://ui.adsabs.harvard.edu/abs/2000AJ....119..800D} {119, 800}

\bibitem[\protect\citeauthoryear{{Fragkoudi} et~al.,}{{Fragkoudi}
  et~al.}{2019}]{fragkoudi2019ridges}
{Fragkoudi} F.,  et~al., 2019, \mn@doi [\mnras] {10.1093/mnras/stz1875}, \href
  {https://ui.adsabs.harvard.edu/abs/2019MNRAS.488.3324F} {488, 3324}

\bibitem[\protect\citeauthoryear{{Friske} \& {Sch{\"o}nrich}}{{Friske} \&
  {Sch{\"o}nrich}}{2019}]{friske2019more}
{Friske} J. K.~S.,  {Sch{\"o}nrich} R.,  2019, \mn@doi [\mnras]
  {10.1093/mnras/stz2951}, \href
  {https://ui.adsabs.harvard.edu/abs/2019MNRAS.490.5414F} {490, 5414}

\bibitem[\protect\citeauthoryear{{Gaia Collaboration} et~al.,}{{Gaia
  Collaboration} et~al.}{2018}]{GaiaDR2GaiaCollaboration}
{Gaia Collaboration} et~al., 2018, \mn@doi [\aap]
  {10.1051/0004-6361/201833051}, \href
  {http://adsabs.harvard.edu/abs/2018A%26A...616A...1G} {616, A1}

\bibitem[\protect\citeauthoryear{{Ghafourian}, {Roshan}  \&
  {Abbassi}}{{Ghafourian} et~al.}{2020}]{Ghafourian2020Modified}
{Ghafourian} N.,  {Roshan} M.,   {Abbassi} S.,  2020, \mn@doi [\apj]
  {10.3847/1538-4357/ab8c4b}, \href
  {https://ui.adsabs.harvard.edu/abs/2020ApJ...895...13G} {895, 13}

\bibitem[\protect\citeauthoryear{{Goodman}}{{Goodman}}{2000}]{Goodman2000Repulsive}
{Goodman} J.,  2000, \mn@doi [\na] {10.1016/S1384-1076(00)00015-4}, \href
  {https://ui.adsabs.harvard.edu/abs/2000NewA....5..103G} {5, 103}

\bibitem[\protect\citeauthoryear{{Gravity Collaboration} et~al.,}{{Gravity
  Collaboration} et~al.}{2019}]{Gravity2019geometric}
{Gravity Collaboration} et~al., 2019, \mn@doi [\aap]
  {10.1051/0004-6361/201935656}, \href
  {https://ui.adsabs.harvard.edu/abs/2019A&A...625L..10G} {625, L10}

\bibitem[\protect\citeauthoryear{{Grenon}}{{Grenon}}{1972}]{Grenon1972SMR}
{Grenon} M.,  1972, in {Cayrel de Strobel} G.,  {Delplace} A.~M.,  eds, IAU
  Colloq. 17: Age des Etoiles. p.~55

\bibitem[\protect\citeauthoryear{{Grenon}}{{Grenon}}{1999}]{Grenon1999Kinematics}
{Grenon} M.,  1999, in {Spite} M.,  ed., Galaxy Evolution: Connecting the
  Distant Universe with the Local Fossil Record. Kluwer Academic Publishers,
  p.~331

\bibitem[\protect\citeauthoryear{{Halle}, {Di Matteo}, {Haywood}  \&
  {Combes}}{{Halle} et~al.}{2018}]{Halle2018Radial}
{Halle} A.,  {Di Matteo} P.,  {Haywood} M.,   {Combes} F.,  2018, \mn@doi
  [\aap] {10.1051/0004-6361/201832603}, \href
  {https://ui.adsabs.harvard.edu/abs/2018A&A...616A..86H} {616, A86}

\bibitem[\protect\citeauthoryear{{Henrard}}{{Henrard}}{1982}]{Henrard1982Capture}
{Henrard} J.,  1982, \mn@doi [Celestial Mechanics] {10.1007/BF01228946}, \href
  {https://ui.adsabs.harvard.edu/abs/1982CeMec..27....3H} {27, 3}

\bibitem[\protect\citeauthoryear{{Hernquist} \& {Weinberg}}{{Hernquist} \&
  {Weinberg}}{1992}]{hernquist1992bar}
{Hernquist} L.,  {Weinberg} M.~D.,  1992, \mn@doi [\apj] {10.1086/171975},
  \href {https://ui.adsabs.harvard.edu/abs/1992ApJ...400...80H} {400, 80}

\bibitem[\protect\citeauthoryear{{Hu}, {Barkana}  \& {Gruzinov}}{{Hu}
  et~al.}{2000}]{Hu2000Fuzzy}
{Hu} W.,  {Barkana} R.,   {Gruzinov} A.,  2000, \mn@doi [\prl]
  {10.1103/PhysRevLett.85.1158}, \href
  {https://ui.adsabs.harvard.edu/abs/2000PhRvL..85.1158H} {85, 1158}

\bibitem[\protect\citeauthoryear{{Hunt} \& {Bovy}}{{Hunt} \&
  {Bovy}}{2018}]{Hunt2018OUHR}
{Hunt} J. A.~S.,  {Bovy} J.,  2018, \mn@doi [\mnras] {10.1093/mnras/sty921},
  \href {https://ui.adsabs.harvard.edu/abs/2018MNRAS.477.3945H} {477, 3945}

\bibitem[\protect\citeauthoryear{{Hunt}, {Hong}, {Bovy}, {Kawata}  \&
  {Grand}}{{Hunt} et~al.}{2018}]{Hunt2018Transient}
{Hunt} J. A.~S.,  {Hong} J.,  {Bovy} J.,  {Kawata} D.,   {Grand} R. J.~J.,
  2018, \mn@doi [\mnras] {10.1093/mnras/sty2532}, \href
  {https://ui.adsabs.harvard.edu/abs/2018MNRAS.481.3794H} {481, 3794}

\bibitem[\protect\citeauthoryear{{Joshi}}{{Joshi}}{2007}]{Joshi07}
{Joshi} Y.~C.,  2007, \mn@doi [\mnras] {10.1111/j.1365-2966.2007.11831.x},
  \href {https://ui.adsabs.harvard.edu/abs/2007MNRAS.378..768J} {378, 768}

\bibitem[\protect\citeauthoryear{{Katz} et~al.,}{{Katz}
  et~al.}{2019}]{GaiaDR2Katz2019}
{Katz} D.,  et~al., 2019, \mn@doi [\aap] {10.1051/0004-6361/201833273}, \href
  {http://adsabs.harvard.edu/abs/2019A%26A...622A.205K} {622, A205}

\bibitem[\protect\citeauthoryear{{Kushniruk}, {Bensby}, {Feltzing},
  {Sahlholdt}, {Feuillet}  \& {Casagrande}}{{Kushniruk}
  et~al.}{2020}]{Kushniruk2020HR1614}
{Kushniruk} I.,  {Bensby} T.,  {Feltzing} S.,  {Sahlholdt} C.~L.,  {Feuillet}
  D.,   {Casagrande} L.,  2020, \mn@doi [\aap] {10.1051/0004-6361/202037923},
  \href {https://ui.adsabs.harvard.edu/abs/2020A&A...638A.154K} {638, A154}

\bibitem[\protect\citeauthoryear{{Leaman}}{{Leaman}}{2012}]{Leaman2012Insights}
{Leaman} R.,  2012, \mn@doi [\aj] {10.1088/0004-6256/144/6/183}, \href
  {https://ui.adsabs.harvard.edu/abs/2012AJ....144..183L} {144, 183}

\bibitem[\protect\citeauthoryear{{Lichtenberg} \& {Lieberman}}{{Lichtenberg} \&
  {Lieberman}}{1992}]{lichtenberg1992regular}
{Lichtenberg} A.,  {Lieberman} M.,  1992, {Regular and Chaotic Dynamics}.
Springer-Verlag

\bibitem[\protect\citeauthoryear{{Luck}}{{Luck}}{2018}]{Luck2018Cepheid}
{Luck} R.~E.,  2018, \mn@doi [\aj] {10.3847/1538-3881/aadcac}, \href
  {https://ui.adsabs.harvard.edu/abs/2018AJ....156..171L} {156, 171}

\bibitem[\protect\citeauthoryear{{Martinez-Valpuesta}, {Shlosman}  \&
  {Heller}}{{Martinez-Valpuesta} et~al.}{2006}]{Martinez2006Evolution}
{Martinez-Valpuesta} I.,  {Shlosman} I.,   {Heller} C.,  2006, \mn@doi [\apj]
  {10.1086/498338}, \href
  {https://ui.adsabs.harvard.edu/abs/2006ApJ...637..214M} {637, 214}

\bibitem[\protect\citeauthoryear{{McMillan}}{{McMillan}}{2017}]{McMillan2017mass}
{McMillan} P.~J.,  2017, \mn@doi [\mnras] {10.1093/mnras/stw2759}, \href
  {https://ui.adsabs.harvard.edu/abs/2017MNRAS.465...76M} {465, 76}

\bibitem[\protect\citeauthoryear{{Monari}, {Famaey}, {Siebert}, {Wegg}  \&
  {Gerhard}}{{Monari} et~al.}{2019}]{Monari2019signatures}
{Monari} G.,  {Famaey} B.,  {Siebert} A.,  {Wegg} C.,   {Gerhard} O.,  2019,
  \mn@doi [\aap] {10.1051/0004-6361/201834820}, \href
  {https://ui.adsabs.harvard.edu/abs/2019A&A...626A..41M} {626, A41}

\bibitem[\protect\citeauthoryear{{Netopil}, {Paunzen}, {Heiter}  \&
  {Soubiran}}{{Netopil} et~al.}{2016}]{Netopil2016opencluster}
{Netopil} M.,  {Paunzen} E.,  {Heiter} U.,   {Soubiran} C.,  2016, \mn@doi
  [\aap] {10.1051/0004-6361/201526370}, \href
  {https://ui.adsabs.harvard.edu/abs/2016A&A...585A.150N} {585, A150}

\bibitem[\protect\citeauthoryear{{P{\'e}rez-Villegas}, {Portail}, {Wegg}  \&
  {Gerhard}}{{P{\'e}rez-Villegas} et~al.}{2017}]{perez2017revisiting}
{P{\'e}rez-Villegas} A.,  {Portail} M.,  {Wegg} C.,   {Gerhard} O.,  2017,
  \mn@doi [\apjl] {10.3847/2041-8213/aa6c26}, \href
  {https://ui.adsabs.harvard.edu/abs/2017ApJ...840L...2P} {840, L2}

\bibitem[\protect\citeauthoryear{{Portail}, {Gerhard}, {Wegg}  \&
  {Ness}}{{Portail} et~al.}{2017}]{Portail2017Dynamical}
{Portail} M.,  {Gerhard} O.,  {Wegg} C.,   {Ness} M.,  2017, \mn@doi [\mnras]
  {10.1093/mnras/stw2819}, \href
  {https://ui.adsabs.harvard.edu/abs/2017MNRAS.465.1621P} {465, 1621}

\bibitem[\protect\citeauthoryear{Quinn \& Rand}{Quinn \&
  Rand}{1995}]{Quinn1995perturbation}
Quinn D.~D.,  Rand R.~H.,  1995, Smart Structures, Nonlinear Dynamics, and
  Control, pp 226--246

\bibitem[\protect\citeauthoryear{{Reid} \& {Brunthaler}}{{Reid} \&
  {Brunthaler}}{2020}]{Reid2020SagittariusA}
{Reid} M.~J.,  {Brunthaler} A.,  2020, \mn@doi [\apj]
  {10.3847/1538-4357/ab76cd}, \href
  {https://ui.adsabs.harvard.edu/abs/2020ApJ...892...39R} {892, 39}

\bibitem[\protect\citeauthoryear{{Reid} et~al.,}{{Reid}
  et~al.}{2019}]{Reid2019Trigonometric}
{Reid} M.~J.,  et~al., 2019, \mn@doi [\apj] {10.3847/1538-4357/ab4a11}, \href
  {https://ui.adsabs.harvard.edu/abs/2019ApJ...885..131R} {885, 131}

\bibitem[\protect\citeauthoryear{{Sanders}, {Smith}  \& {Evans}}{{Sanders}
  et~al.}{2019}]{Sanders2019pattern}
{Sanders} J.~L.,  {Smith} L.,   {Evans} N.~W.,  2019, \mn@doi [\mnras]
  {10.1093/mnras/stz1827}, \href
  {https://ui.adsabs.harvard.edu/abs/2019MNRAS.488.4552S} {488, 4552}

\bibitem[\protect\citeauthoryear{{Sartoretti} et~al.,}{{Sartoretti}
  et~al.}{2018}]{GaiaDR2Sartoretti2018}
{Sartoretti} P.,  et~al., 2018, \mn@doi [\aap] {10.1051/0004-6361/201832836},
  \href {https://ui.adsabs.harvard.edu/abs/2018A&A...616A...6S} {616, A6}

\bibitem[\protect\citeauthoryear{{Sch{\"o}nrich} \& {McMillan}}{{Sch{\"o}nrich}
  \& {McMillan}}{2017}]{Schoenrich2017Understanding}
{Sch{\"o}nrich} R.,  {McMillan} P.~J.,  2017, \mn@doi [\mnras]
  {10.1093/mnras/stx093}, \href
  {https://ui.adsabs.harvard.edu/abs/2017MNRAS.467.1154S} {467, 1154}

\bibitem[\protect\citeauthoryear{{Sch{\"o}nrich}, {Binney}  \&
  {Dehnen}}{{Sch{\"o}nrich} et~al.}{2010}]{SBD}
{Sch{\"o}nrich} R.,  {Binney} J.,   {Dehnen} W.,  2010, \mn@doi [\mnras]
  {10.1111/j.1365-2966.2010.16253.x}, \href
  {https://ui.adsabs.harvard.edu/abs/2010MNRAS.403.1829S} {403, 1829}

\bibitem[\protect\citeauthoryear{{Sch{\"o}nrich}, {McMillan}  \&
  {Eyer}}{{Sch{\"o}nrich} et~al.}{2019}]{Schoenrich2019distance}
{Sch{\"o}nrich} R.,  {McMillan} P.,   {Eyer} L.,  2019, \mn@doi [\mnras]
  {10.1093/mnras/stz1451}, \href
  {https://ui.adsabs.harvard.edu/abs/2019MNRAS.tmp.1390S} {p.~1390}

\bibitem[\protect\citeauthoryear{{Sellwood}}{{Sellwood}}{2008}]{Sellwood2008BarHaloIII}
{Sellwood} J.~A.,  2008, \mn@doi [\apj] {10.1086/586882}, \href
  {https://ui.adsabs.harvard.edu/abs/2008ApJ...679..379S} {679, 379}

\bibitem[\protect\citeauthoryear{{Sormani}, {Binney}  \& {Magorrian}}{{Sormani}
  et~al.}{2015}]{sormani2015gas3}
{Sormani} M.~C.,  {Binney} J.,   {Magorrian} J.,  2015, \mn@doi [\mnras]
  {10.1093/mnras/stv2067}, \href
  {https://ui.adsabs.harvard.edu/abs/2015MNRAS.454.1818S} {454, 1818}

\bibitem[\protect\citeauthoryear{{Spagna}, {Lattanzi}, {Re Fiorentin}  \&
  {Smart}}{{Spagna} et~al.}{2010}]{Spagna2010}
{Spagna} A.,  {Lattanzi} M.~G.,  {Re Fiorentin} P.,   {Smart} R.~L.,  2010,
  \mn@doi [\aap] {10.1051/0004-6361/200913538}, \href
  {https://ui.adsabs.harvard.edu/abs/2010A&A...510L...4S} {510, L4}

\bibitem[\protect\citeauthoryear{{Tremaine} \& {Weinberg}}{{Tremaine} \&
  {Weinberg}}{1984}]{Tremaine1984Dynamical}
{Tremaine} S.,  {Weinberg} M.~D.,  1984, \mn@doi [\mnras]
  {10.1093/mnras/209.4.729}, \href
  {https://ui.adsabs.harvard.edu/abs/1984MNRAS.209..729T} {209, 729}

\bibitem[\protect\citeauthoryear{{Trick}, {Fragkoudi}, {Hunt}, {Mackereth}  \&
  {White}}{{Trick} et~al.}{2021}]{Trick2021Identifying}
{Trick} W.~H.,  {Fragkoudi} F.,  {Hunt} J. A.~S.,  {Mackereth} J.~T.,   {White}
  S. D.~M.,  2021, \mn@doi [\mnras] {10.1093/mnras/staa3317}, \href
  {https://ui.adsabs.harvard.edu/abs/2021MNRAS.500.2645T} {500, 2645}

\bibitem[\protect\citeauthoryear{{Wegg}, {Gerhard}  \& {Portail}}{{Wegg}
  et~al.}{2015}]{wegg2015structure}
{Wegg} C.,  {Gerhard} O.,   {Portail} M.,  2015, \mn@doi [\mnras]
  {10.1093/mnras/stv745}, \href
  {https://ui.adsabs.harvard.edu/abs/2015MNRAS.450.4050W} {450, 4050}

\bibitem[\protect\citeauthoryear{{Weiler}}{{Weiler}}{2018}]{Weiler2018Revised}
{Weiler} M.,  2018, \mn@doi [\aap] {10.1051/0004-6361/201833462}, \href
  {https://ui.adsabs.harvard.edu/abs/2018A&A...617A.138W} {617, A138}

\bibitem[\protect\citeauthoryear{{Weinberg}}{{Weinberg}}{1985}]{weinberg1985evolution}
{Weinberg} M.~D.,  1985, \mn@doi [\mnras] {10.1093/mnras/213.3.451}, \href
  {https://ui.adsabs.harvard.edu/abs/1985MNRAS.213..451W} {213, 451}

\bibitem[\protect\citeauthoryear{{Weinberg}}{{Weinberg}}{2004}]{Weinberg2004Timedependent}
{Weinberg} M.~D.,  2004, arXiv e-prints, \href
  {https://ui.adsabs.harvard.edu/abs/2004astro.ph..4169W} {pp
  astro--ph/0404169}

\bibitem[\protect\citeauthoryear{{Weinberg} \& {Katz}}{{Weinberg} \&
  {Katz}}{2002}]{Weinberg2002BarDriven}
{Weinberg} M.~D.,  {Katz} N.,  2002, \mn@doi [\apj] {10.1086/343847}, \href
  {https://ui.adsabs.harvard.edu/abs/2002ApJ...580..627W} {580, 627}

\bibitem[\protect\citeauthoryear{{Weinberg} \& {Katz}}{{Weinberg} \&
  {Katz}}{2007}]{Weinberg2007BarHaloInteraction}
{Weinberg} M.~D.,  {Katz} N.,  2007, \mn@doi [\mnras]
  {10.1111/j.1365-2966.2006.11306.x}, \href
  {https://ui.adsabs.harvard.edu/abs/2007MNRAS.375..425W} {375, 425}

\makeatother
\end{thebibliography}



\appendix

\section{Model}
\label{sec:app_model}

We model the Galaxy as a Mestel disc perturbed by a decelerating/elongating bar described by a quadrupole term:
\begin{align}
  \Phi(R, \varphi, t) 
  &= \vc^2 \ln\left(R\right) - \frac{A \vc^2}{m} \left[\frac{R}{\RCR(t)}\right]^2 \left[\frac{b + 1}{b + R/\RCR(t)}\right]^5 \nonumber \\
  &\hspace{4mm}\cos m \left[\varphi - \int_0^t dt'~\Omegap(t') \right],
  \label{eq:bar_potential_amp}
\end{align}
where $\vc = 235 \kms$ is the circular speed \citep{Reid2019Trigonometric}. The parameter $A$ describes the strength of the bar (ratio between the maximum azimuthal force by the bar and the radial force due to the unperturbed potential at $\RCR$), and $b$ is the ratio of the bar scale length to $\RCR$. The bar's pattern speed $\Omegap(t)$ is modelled to decrease inversely proportional to time corresponding to a linear increase in the corotation radius $\RCR(t) \equiv \vc / \Omegap(t)$. The slowing rate of the bar is then conveniently described by a constant, dimensionless parameter $\eta \equiv -\dOmegap/\Omegap^2 = \dot{R}_{\rm CR}/\vc$. Details on our model are described in \cite{Chiba2020ResonanceSweeping}.

\section{Calculation of libration action.}
\label{sec:app_librationaction}

The motion of orbits trapped and dragged by a slowing bar is described to first order by a differential equation that represents a pendulum subject to a constant torque \citep{Tremaine1984Dynamical,Chiba2020ResonanceSweeping}:
\begin{align}
  \ddthetas + \omega^2 \left( \sin \thetas - \frac{\eta}{A} \right) = 0,
  \label{eq:thetas_diff_eq}
\end{align}
where $\omega^2 \equiv - G \Psi$, $G \equiv \frac{\partial^2 H_0}{\partial \Js^2}$, and $\Psi$ is the Fourier coefficient of the bar potential expanded in slow-fast angle variables \citep{Chiba2020ResonanceSweeping}. Both $G$ and $\Psi$ are evaluated at the centre of the moving resonance $\Jsres(t)$. If the bar decelerates slowly such that the temporal change in $\omega(t)$ is negligibly slow compared to the evolution of $\thetas$, we may write
\begin{align}
  \Ep = \frac{1}{2} \dthetas^2 + V(\thetas) ~,~~~ V(\thetas) = \omega^2 \left( - \cos \thetas - \frac{\eta}{A} \thetas \right),
  \label{eq:Ep}
\end{align}
where $\dthetas = G \Delta$ and $\Delta \equiv \Js - \Jsres$. As in the analogous case of a harmonic oscillator, $\Ep$ is not conserved under adiabatic/slow changes in $\omega$, while the associated action of libration is approximately conserved:
\begin{align}
  \Jl = \int_C \frac{d\thetas}{2\pi} \left( \Delta_{+} - \Delta_{-} \right) 
  = \frac{1}{|G|} \int_C \frac{d\thetas}{\pi} \sqrt{2 \left[ \Ep - V(\thetas) \right]},
  \label{eq:Jl}
\end{align}
where $\Delta_{\pm}$ are the roots of the quadratic \eqref{eq:Ep} and the integral $C$ runs from $-\pi$ to $\pi$ wherever $\Delta_{\pm}$ is real. The maximum libration action is given by the minimum $\Ep$ necessary to reach the crest of the potential (i.e. $\frac{dV}{d\thetas}\big|_{\thetasep} = 0$):
\begin{align}
  \Jlsep &= \sqrt{\frac{\Psi}{|G|}} \int_C \frac{d\thetas}{\pi} \sqrt{2\left[\cos\thetas - \cos\thetassep + \frac{\eta}{A}\left(\thetas - \thetassep\right)\right]}
  \label{eq:Jlsep}
\end{align}
where $\thetassep = \sin^{-1} \left(\eta/A\right), ~ \pi/2 \leq \thetassep \leq \pi$. In the limit of epicycle approximation, $\Psi$ and $G$ scale as $\Psi \sim A \vc^2$ and $G \sim 1/\RCR^2 \sim \Omegap^2/\vc^2$ (see Appendix of \citealt{Chiba2020ResonanceSweeping}), so $\Jlsep \sim \sqrt{A} \vc^2 / \Omegap$ which qualitatively explains the behaviour presented in Fig.~\ref{fig:wp_Jlsep}. As shown in Fig.~\ref{fig:simulation_slowAA}, \eqref{eq:Jlsep} successfully marks the phase-space boundary of trapped orbits integrated numerically. We note however that this closed curve is not strictly a separatrix. Since the parameter $\omega$ of the pendulum equation (\ref{eq:thetas_diff_eq}) is time-dependent, the separatrix near the saddle point $\thetassep$ is in fact broken (not closed) allowing orbits to enter or leave the resonance therefrom \citep{Quinn1995perturbation}. In this paper, we will nevertheless refer to the phase curve drawn by \eqref{eq:Jlsep} as `separatrix' since it marks the approximate phase-space area of orbits currently trapped in resonance. The separatrix of a resting resonance ($\eta=0$, black dot-dashed) is also drawn in Fig.~\ref{fig:simulation_slowAA} for comparison. The deceleration of the bar has two notable consequences: the volume of resonant phase space shrinks, and the centre of the resonance shifts towards positive $\thetas$; i.e. trapped orbits are azimuthally tilted when seen in the bar's rotating frame \citep[][figure 16]{Chiba2020ResonanceSweeping} which is simply the consequence of the Euler force. Fortunately, the Sun is rotating ahead of the corotating orbits, so the reduction of resonant volume in the Solar neighbourhood is relatively small (Fig.~\ref{fig:simulation_slowAA} right hand panel). \cite{Chiba2020ResonanceSweeping} gave estimates on the slowing rate as $\eta = 0.0036 \pm 0.0011$ by quantifying the asymmetry of the Hercules. In this paper, however, we use the libration action evaluated in a fixed pattern speed as default since the bar's slowing rate contains large uncertainty propagated from the uncertainty in the bar strength. The effect of bar deceleration on the estimation of the pattern speed is examined and reported at the end of section \ref{sec:pattern_speed}. We also checked that the first and second order terms of the Taylor expansion of $\Psi$ around the resonance, which we have neglected, barely affect the estimation of the pattern speed.

\section{Capture probability}
\label{app:capture_probability}

\begin{figure}
  \begin{center}
    \includegraphics[width=8.5cm]{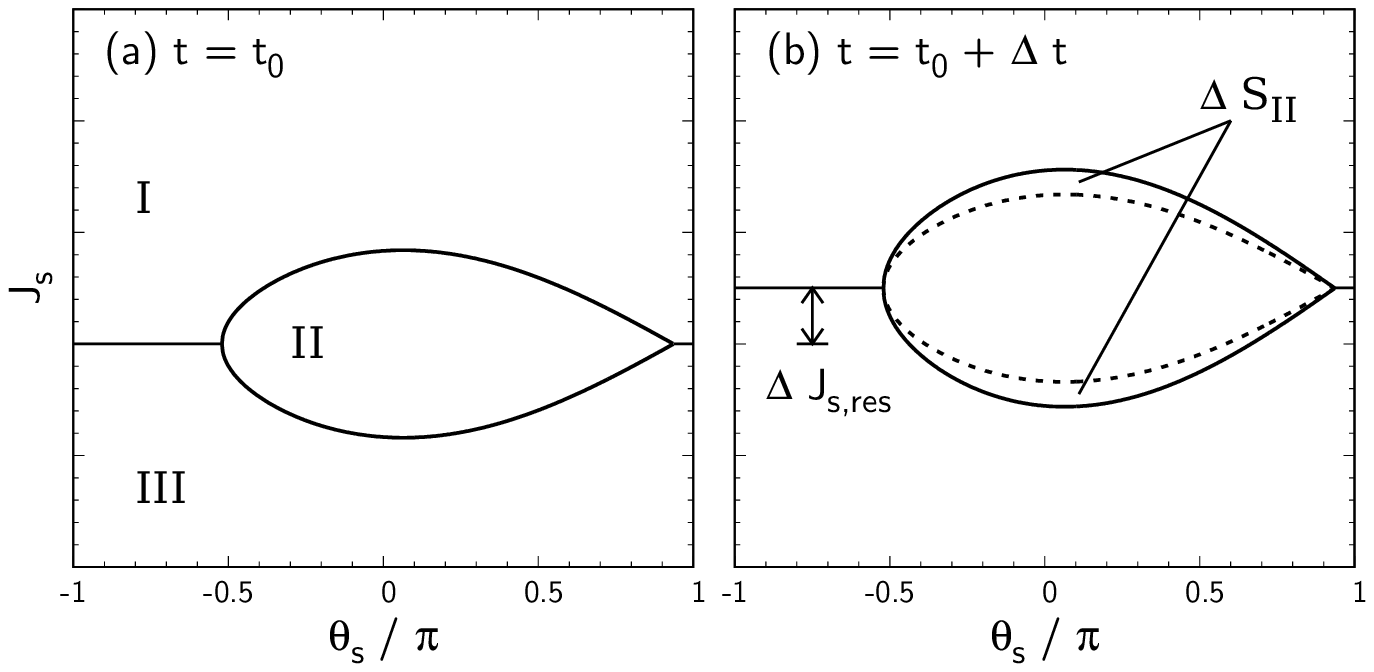}
    \caption{Schematic drawing of the evolution of a resonance. The phase space is split into three regime: the upper circulation regime (I), the libration regime (II), and the lower circulation regime (III). As time $\Delta t$ passes (from left panel to right panel), the resonance moves up by $\Delta \Jsres$ and grows by $\Delta S_{\rm II} = 2 \pi \Delta \Jlsep$. The time variation of the phase-space area of the respective regions determines the capture rate (equation \ref{eq:capture_probability}).}
    \label{fig:capture_diagram}
  \end{center}
\end{figure}

\begin{figure}
  \begin{center}
    \includegraphics[width=8.5cm]{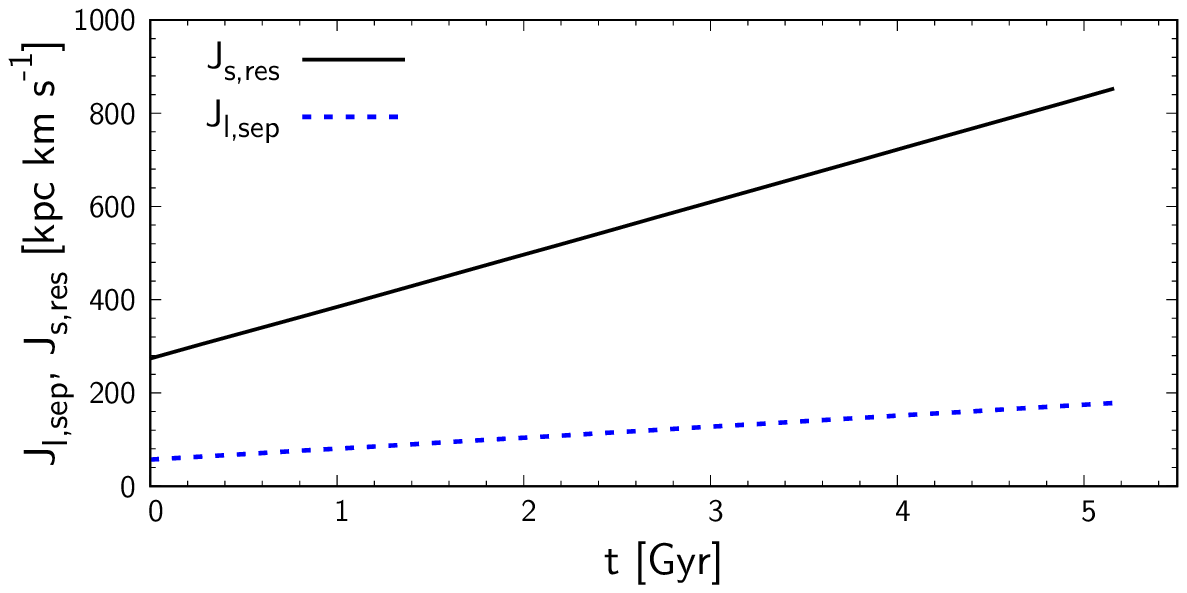}
    \includegraphics[width=8.5cm]{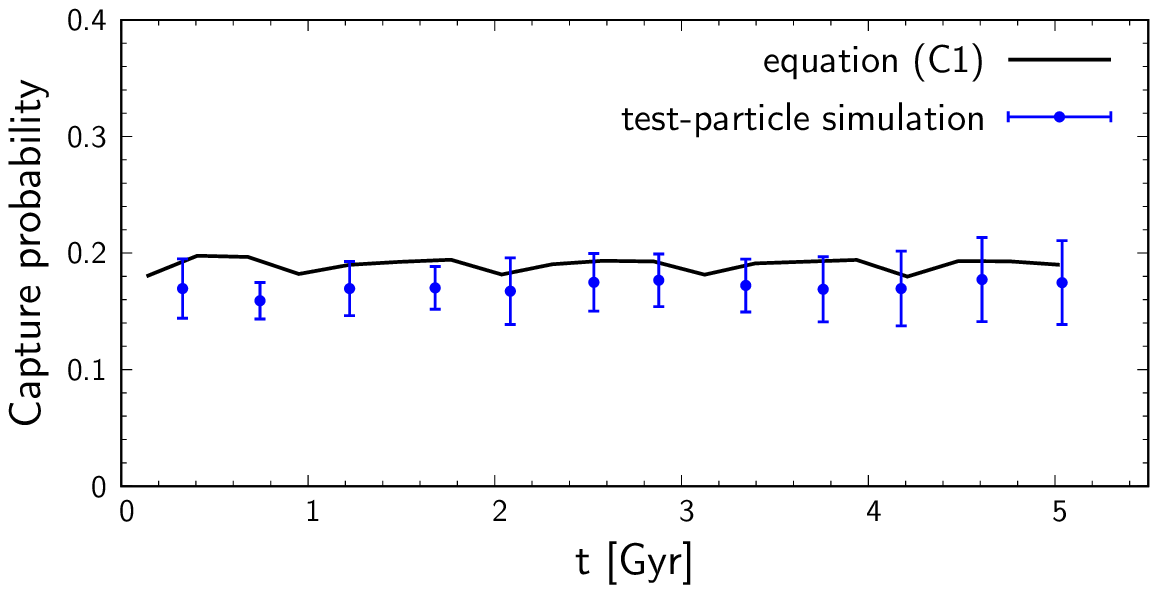}
    \caption{Upper panel: The time evolution of the CR's location $\Jsres$ and its volume $\Jlsep$. Lower panel: The angle-averaged capture probability calculated analytically (black) and numerically (blue).}
    \label{fig:capture_t}
  \end{center}
\end{figure}

Here we report on the capture rate of our slowing bar. Figure~\ref{fig:capture_diagram} depicts the evolution of the contracted resonance with time (from left to right). As the resonance advances in $\Js$, the area of resonance grows (section \ref{subsec:increase_in_resonant_volume}), so a fraction of stars above the resonance (region I) that encounter the separatrix may either be captured into the resonance (region II) or transfer to the lower side (region III) depending on their incident phase. If we assume that trapped orbits always remain trapped as the resonance moves and grows (i.e. if the area enclosed by the dotted curve in Fig.~\ref{fig:capture_diagram} (b) is comprised of stars that occupied region II in Fig.~\ref{fig:capture_diagram} (a)), the capture probability $P_{\rm I \rightarrow \rm II}$ averaged over the phase can be calculated from the rate of change in the phase-space area of each region \citep[][problem 3.43]{Henrard1982Capture,Collett1997Capture,binney2008galactic}:
\begin{align}
  P_{\rm I \rightarrow \rm II} = \fracL{\frac{\drm S_{\rm II}}{\drm t}}{-\frac{\drm S_{\rm I}}{\drm t}},~~~~
  P_{\rm I \rightarrow \rm III} = \fracL{\frac{\drm S_{\rm III}}{\drm t}}{-\frac{\drm S_{\rm I}}{\drm t}},~~~~
  P_{\rm I \rightarrow \rm II} + P_{\rm I \rightarrow \rm III} = 1,
  \label{eq:capture_probability}
\end{align}
where $S_{\rm I,II,III}$ are the phase-space area of region I, II and III respectively which change according to
\begin{align}
  &\frac{\drm S_{\rm I}}{\drm t} = - \frac{1}{2} \frac{\drm S_{\rm II}}{\drm t} - 2 \pi \frac{\drm \Jsres}{\drm t}, \\
  &\frac{\drm S_{\rm II}}{\drm t} = 2 \pi \frac{\drm \Jlsep}{\drm t}, \\
  &\frac{\drm S_{\rm III}}{\drm t} = - \frac{1}{2} \frac{\drm S_{\rm II}}{\drm t} + 2 \pi \frac{\drm \Jsres}{\drm t},
  \label{eq:dSdt}
\end{align}
and are conserved in total
\begin{align}
  \frac{\drm S_{\rm I}}{\drm t} + \frac{\drm S_{\rm II}}{\drm t} + \frac{\drm S_{\rm III}}{\drm t} = 0.
  \label{eq:conservation_of_phasespace}
\end{align}
The capture probability is determined by two factors: how fast the resonance sweeps $\drm \Jsres / \drm t$ and how fast it grows in volume $\drm \Jlsep / \drm t$. Figure~\ref{fig:capture_t} top panel shows the time evolution of $\Jsres$ and $\Jlsep$ at the CR of our slowing/elongating bar model. As in the simulation shown in Fig.~\ref{fig:simulation_slowAA}, the bar amplitude $A$ is kept constant and the fast action is $\Jf = 10 \kpckms$. Since the corotation radius $\RCR$ of our bar is modeled to expand linearly with time (Appendix \ref{sec:app_model}), both $\Jsres$ and $\Jlsep$ are linear as they scale according to $\Jsres \sim \vc \RCR, ~\Jlsep \sim \sqrt{\Psi/|G|} \sim \sqrt{A} \vc \RCR$. Consequently, the capture probability, shown in the bottom panel of Fig.~\ref{fig:capture_t} (black curve), is roughly constant around 0.2. We also calculated the capture rate using test-particle simulation where we place $10^4$ stars with identical initial actions just above the resonance ($J_{\rm s0} = \zeta \Jsres,~ \zeta>1$) but with random angles, and judge capture if $\Js$ increase by a factor of more than 1.2 from the initial value (see \citealt{Chiba2020ResonanceSweeping} section 4.2 for detail). It is self-evident that our measurement is affected by the higher-order resonances which pass the stars before the CR. These resonances take a comparably small phase-space volume that they are still able to temporarily capture and sweep some stars as well as they might bias the angle distribution of stars interacting with the CR. We ascertained that the capture probability to the CR indeed depends at the $20\%$ level on the chosen initial position parametrized by $\zeta$. Thus, we conduct the measurement with $\zeta=1.15,1.20,1.25,1.30$ and take the mean value. The numerical result (blue), which is plotted at the time when the resonance passes $J_{\rm s0}$, is slightly lower than the analytical estimation (black) but reassures the qualitatively behaviour. The overestimation of our analytical approach is most likely due to the assumption that all trapped stars remain trapped as the resonance moves/grows which is invalid at the separatrix where the libration period diverges and thus allows orbits to escape from the moving resonance despite the growth in volume.

\section{Uncertainty in libration action.}
\label{sec:uncertainty_in_Jl}

\begin{figure}
  \begin{center}
    \includegraphics[width=8.5cm]{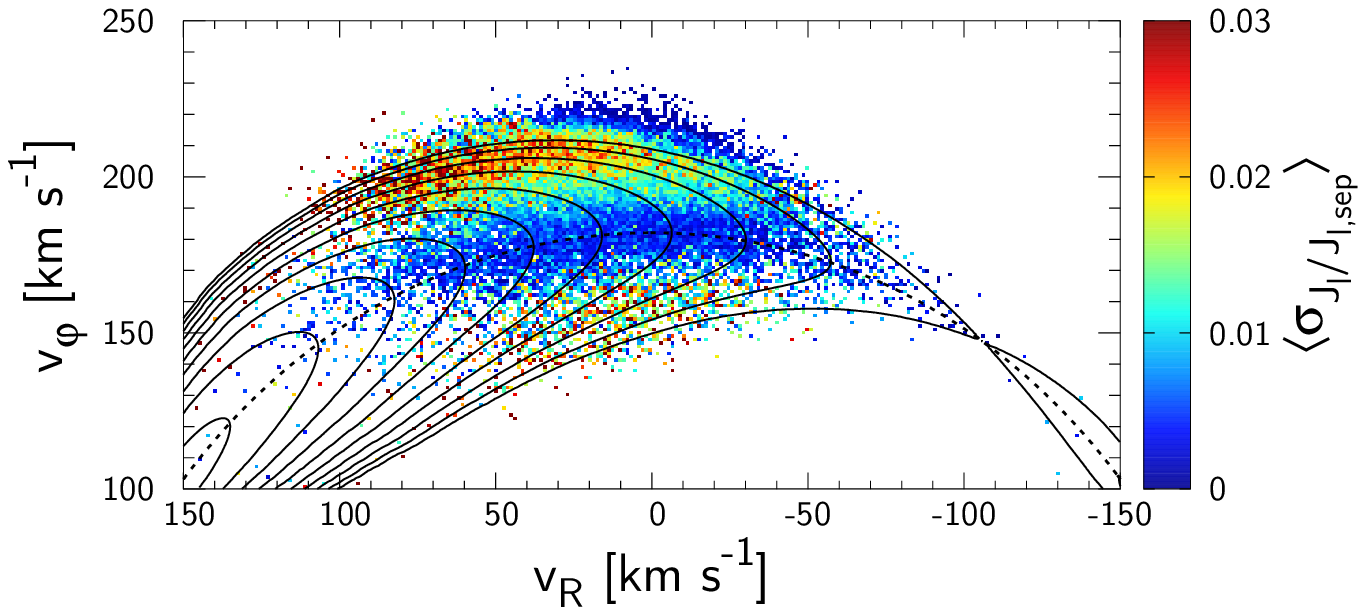}
    \includegraphics[width=8.5cm]{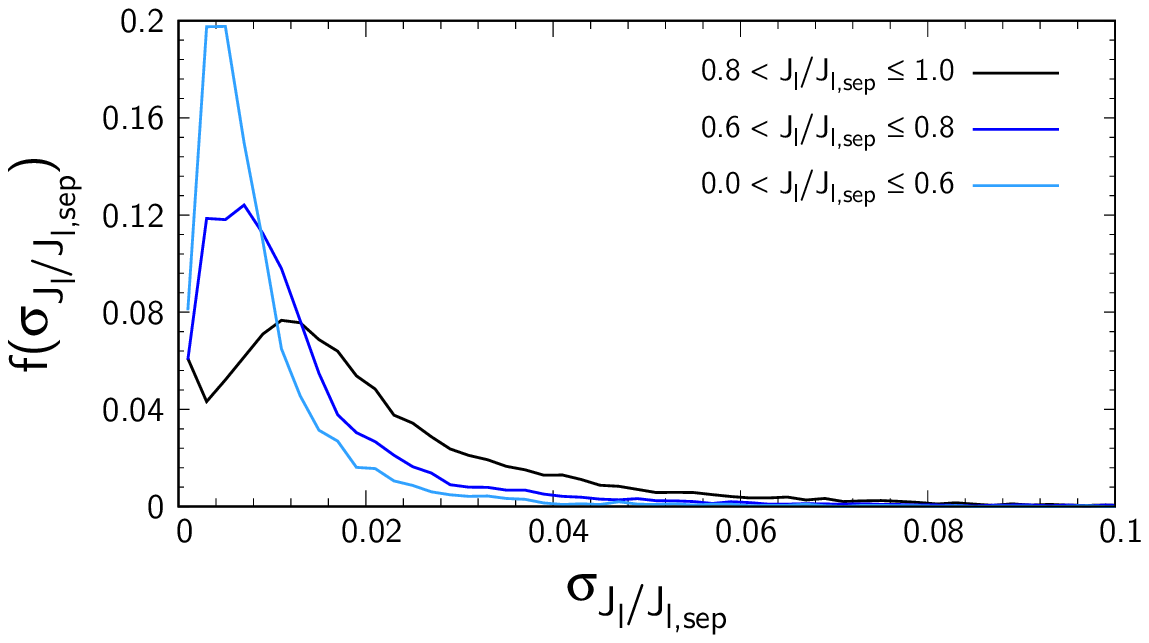}
    \caption{Upper panel: Mean uncertainty of $\Jln$ measurement due to \textit{Gaia} errors. The uncertainty follows the gradient of $\Jln$. Lower panel: The distribution of uncertainty in $\Jln$. For the majority of stars, the uncertainty is of order 0.01 which is sufficiently smaller than the range of $\Jln$ over which we measure the metallicity trend.}
    \label{fig:Jln_err_GaiaDR2}
  \end{center}
\end{figure}

It is important to check that the uncertainty of the libration action $\Jln \equiv \Jl/\Jlsep$ arising from the uncertainties in the \textit{Gaia} data is sufficiently smaller than the scale of $\Jln$ at which we are looking. To quantify the uncertainty in $\Jln$, we prescribe a Gaussian distribution for \textit{Gaia} parallax $p$, proper motions $\mul,\mub$, and line-of-sight velocity $\vlos$ using the reported errors, and estimate the uncertainty by Monte Carlo method with 1000 realization for each star. Figure~\ref{fig:Jln_err_GaiaDR2} upper panel plots the mean uncertainty of $\Jln$ over the velocity space. \textit{Gaia}'s observational errors enter velocity space almost linearly, so the uncertainty in $\Jln$ is largest where the gradient of $\Jln$ in velocity space is steepest. Since the contours of $\Jln$ are calculated using the position of the Sun, some local stars relatively far from the Sun (at most $0.3 \kpc$) appear beyond the separatrix. Figure~\ref{fig:Jln_err_GaiaDR2} lower panel shows the distribution of the uncertainty for three regions of the resonance: the inner region ($0 < \Jln < 0.6$, light blue), the intermediate region ($0.6 < \Jln < 0.8$, blue), and near the separatrix ($0.8 < \Jln < 1$, black). The uncertainty becomes larger towards the separatrix, although for the vast majority of stars it is only a few per cent, so more than an order of magnitude smaller than the range in $\Jln$ across which we have shown the monotonic metallicity increase.

\section{Model of local metallicity.}
\label{sec:app_metallicity_v_model}

\begin{figure*}
  \begin{center}
    \setlength\columnsep{0pt}
    \begin{multicols}{2}
      \includegraphics[width=8.5cm]{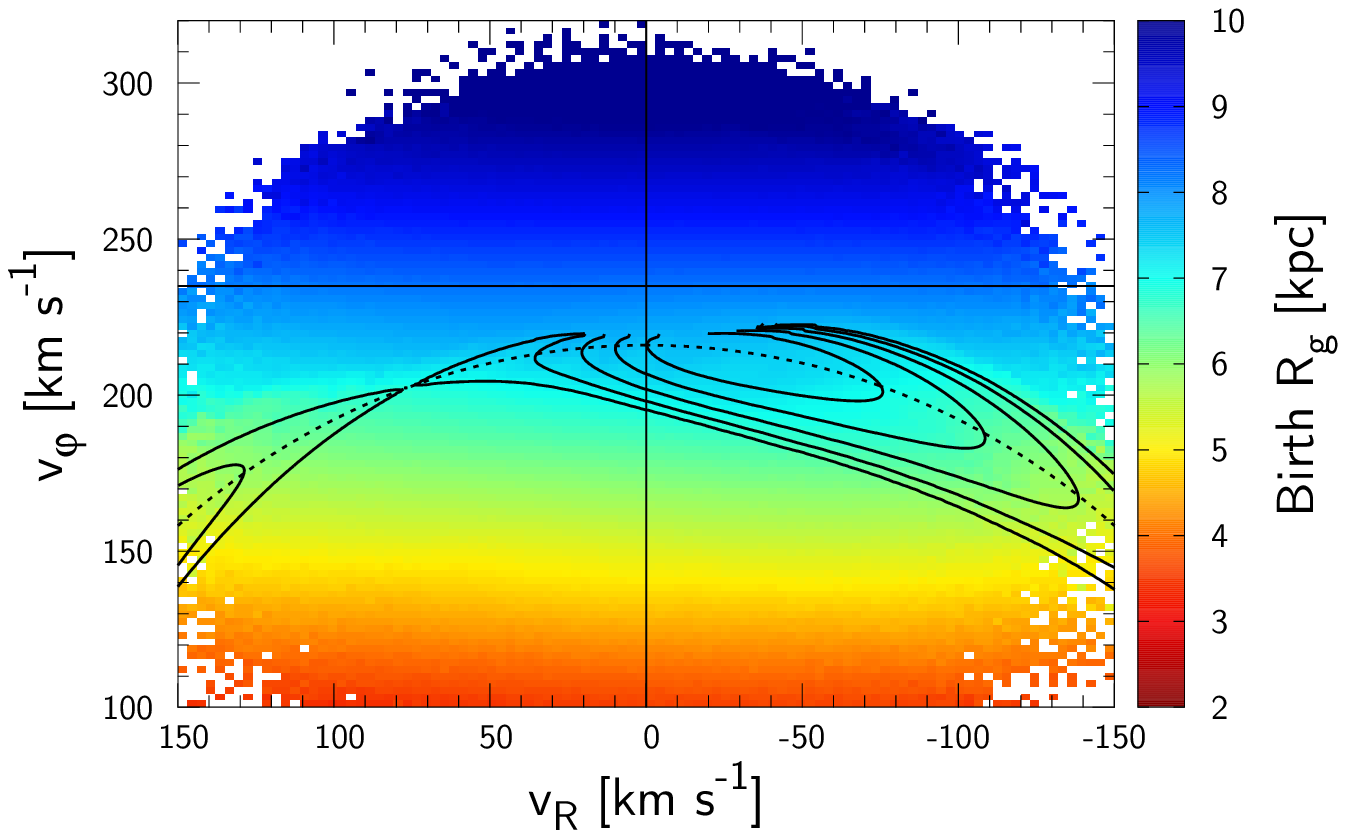}
      \newpage
      \includegraphics[width=8.5cm]{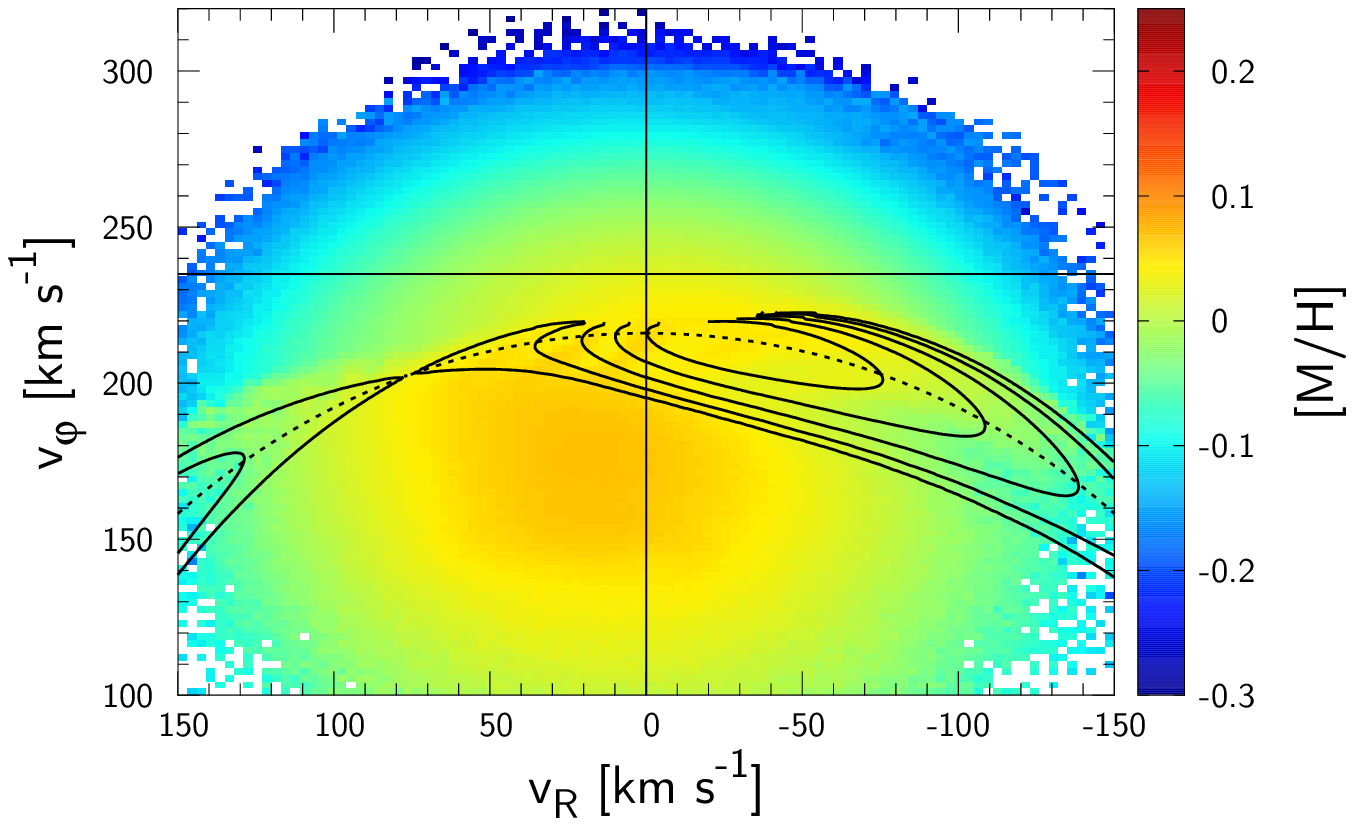}
    \end{multicols}
    (a)~Constant fast bar $(\eta=0,~\Omegap = 53 \kmskpc = 1.85 \Omega_0)$.
    \begin{multicols}{2}
      \includegraphics[width=8.5cm]{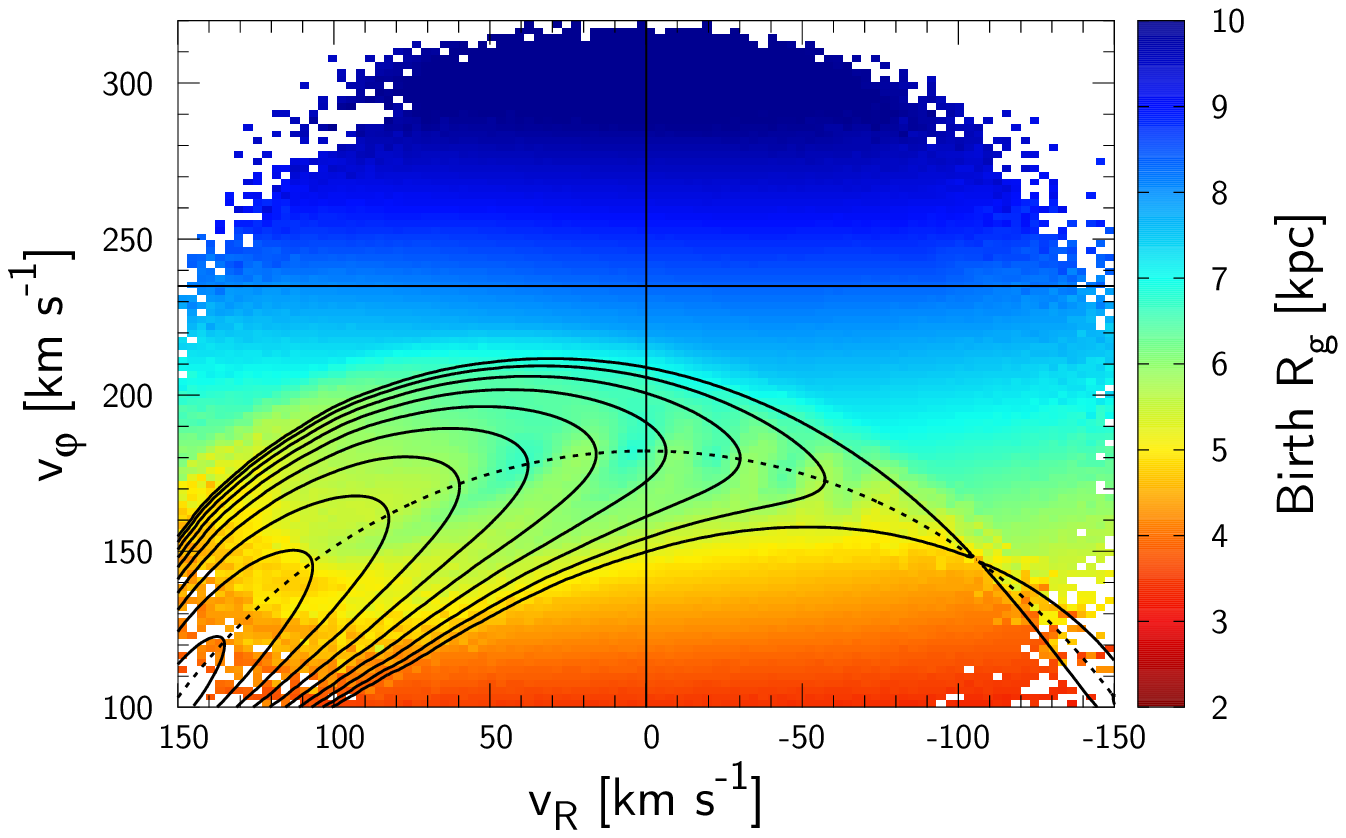}
      \newpage
      \includegraphics[width=8.5cm]{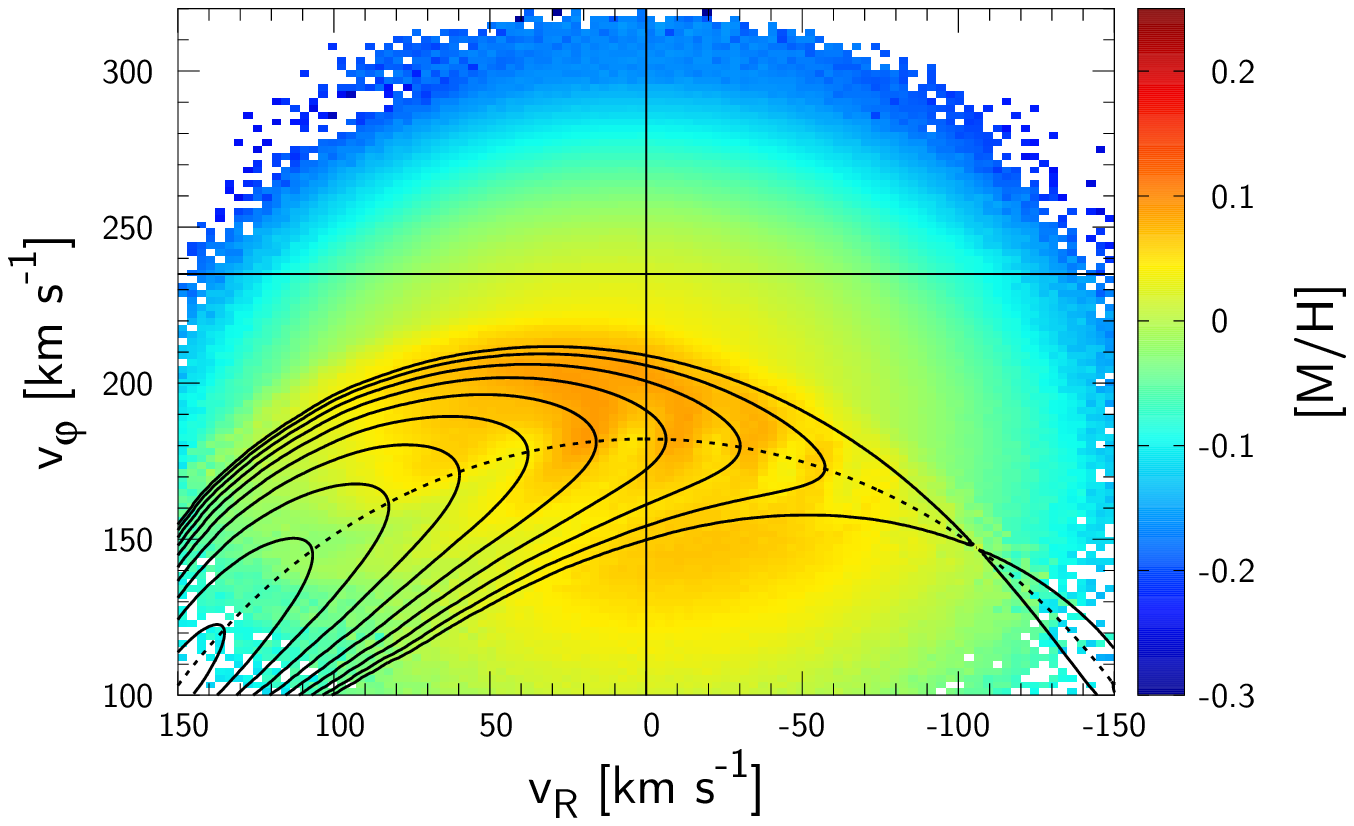}
    \end{multicols}
    (b)~Constant slow bar $(\eta=0,~\Omegap = 35 \kmskpc = 1.22 \Omega_0)$.
    \begin{multicols}{2}
      \includegraphics[width=8.5cm]{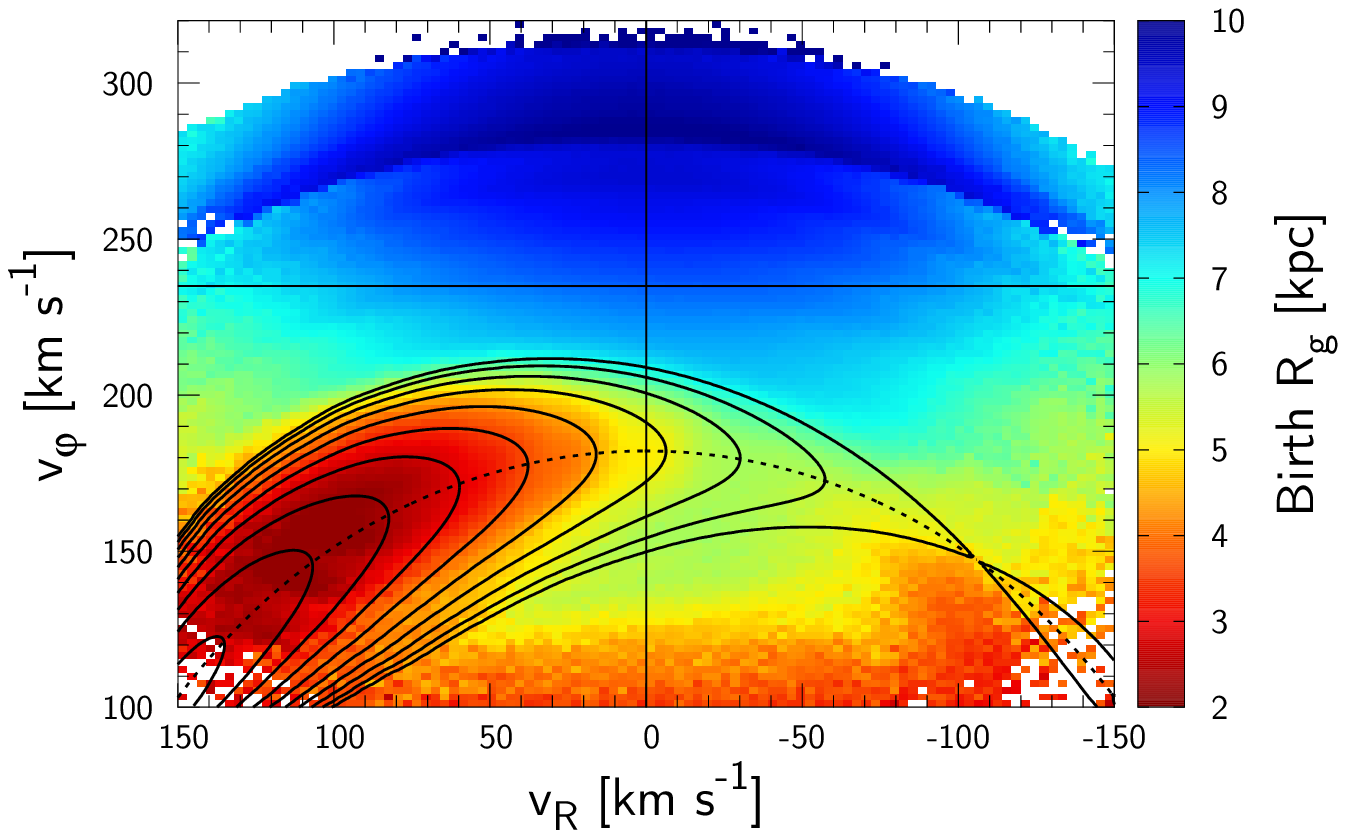}
      \newpage
      \includegraphics[width=8.5cm]{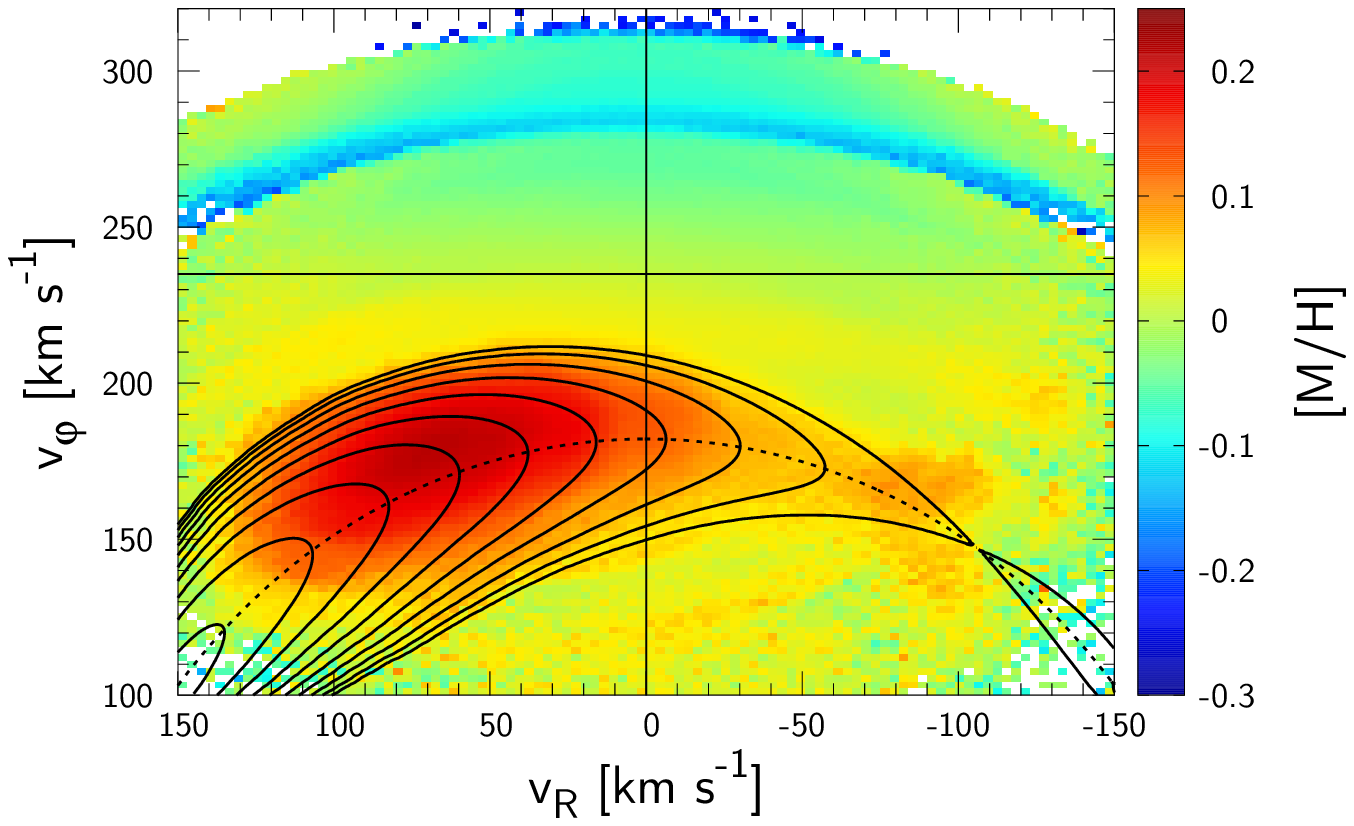}
    \end{multicols}
    (c)~Decelerating slow bar $(\eta=0.004,~\Omegap = 35 \kmskpc = 1.22 \Omega_0)$.
    \caption{The birth guiding radius (left column) and the estimated local metallicity (right column) for three bar models. Top panels: a constant fast bar $(\Omegap = 53 \kmskpc = 1.85 \Omega_0)$. Middle panels: a constant slow bar $(\Omegap = 35 \kmskpc = 1.22 \Omega_0)$. Bottom panels: a decelerating slow bar.}
    \label{fig:vRvphiLz0_vRvphiMH}
  \end{center}
\end{figure*}

Modelling the local metallicity distribution is an involved task requiring a chemo-dynamical model that deals with the age-dispersion relation, the variation of the radial metallicity gradient with age, and radial mixing of stars, all of which is non-trivial and is thus beyond the scope of this paper. Here we just provide a rough guidance on how the different bar models would translate into an observable metallicity distribution.

As evident from the data, both the age-metallicity and the age-velocity dispersion relationships combine into a negative metallicity gradient towards larger radial action which is superpositioned to the $\Jphi$ dependence resulting from the radial metallicity gradient. We fit this metallicity-$\JR$ relationship linearly to roughly cover this effect: $\drm \MH / \drm \JR = -0.00078 \dex \kpc^{-1}\,{\rm km}^{-1}\,{\rm s}$. This relation is then applied to the original $\JR$ of the test particles together with the radial metallicity gradient $-0.05 \dexkpc$ \citep{Luck2018Cepheid} applied to the birth $\Rg$ (i.e. original $\Lz$).

Fig.~\ref{fig:vRvphiLz0_vRvphiMH} displays the birth guiding radius (left column) and the corresponding metallicity (right column) of three different bar models. The top panels show the previously favored fast bar model with pattern speed $\Omegap = 53 \kmskpc$ \citep{Dehnen1999Pattern}. The OLR is located above the Hercules stream and the non-resonant $x_2$ orbits below the OLR constitute the Hercules. The contours of $\Jln$ are broken at small $\JR$ due to the failure of the pendulum formalism at the Lindblad resonances where trapped stars are modeled to librate down to negative $\JR$, although this problem could be resolved by appropriate coordinate transformation \citep{Binney2020Lindblad}. As demonstrated by \cite{Antoja2017RAVE}, the left panel shows that the non-trapped stars at the position of Hercules originate from smaller radii than the OLR stars with similar $\vphi$. Note that \cite{Antoja2017RAVE} plotted the current mean radius of the stars whereas our plot depicts the original guiding radius which is the relevant quantity for assessing the metallicity gradient. As can be seen from our plots, the difference in the original $\Rg$ is too small such that, when the simple metallicity-age-dispersion relation is applied, the metallicity at Hercules is only $0.05 \dex$ larger than that at the LSR while the data shows a difference of more than $\Delta \MH > 0.2 \dex$ (Fig.~\ref{fig:v_kin_MH}).

The middle panels show the slow bar model with constant pattern speed. The Hercules is now associated with orbits trapped in the CR which have a larger range of radial oscillation compared to non-trapped orbits, resulting in a slightly higher metallicity than the fast bar model. However, the predicted metallicity remains below the level of the data. Note the stripes along the contours of libration action arising from the incomplete phase mixing inside the resonance (even though we have run the simulation for $12 \Gyr$).

\begin{figure}
  \begin{center}
    \includegraphics[width=8.5cm]{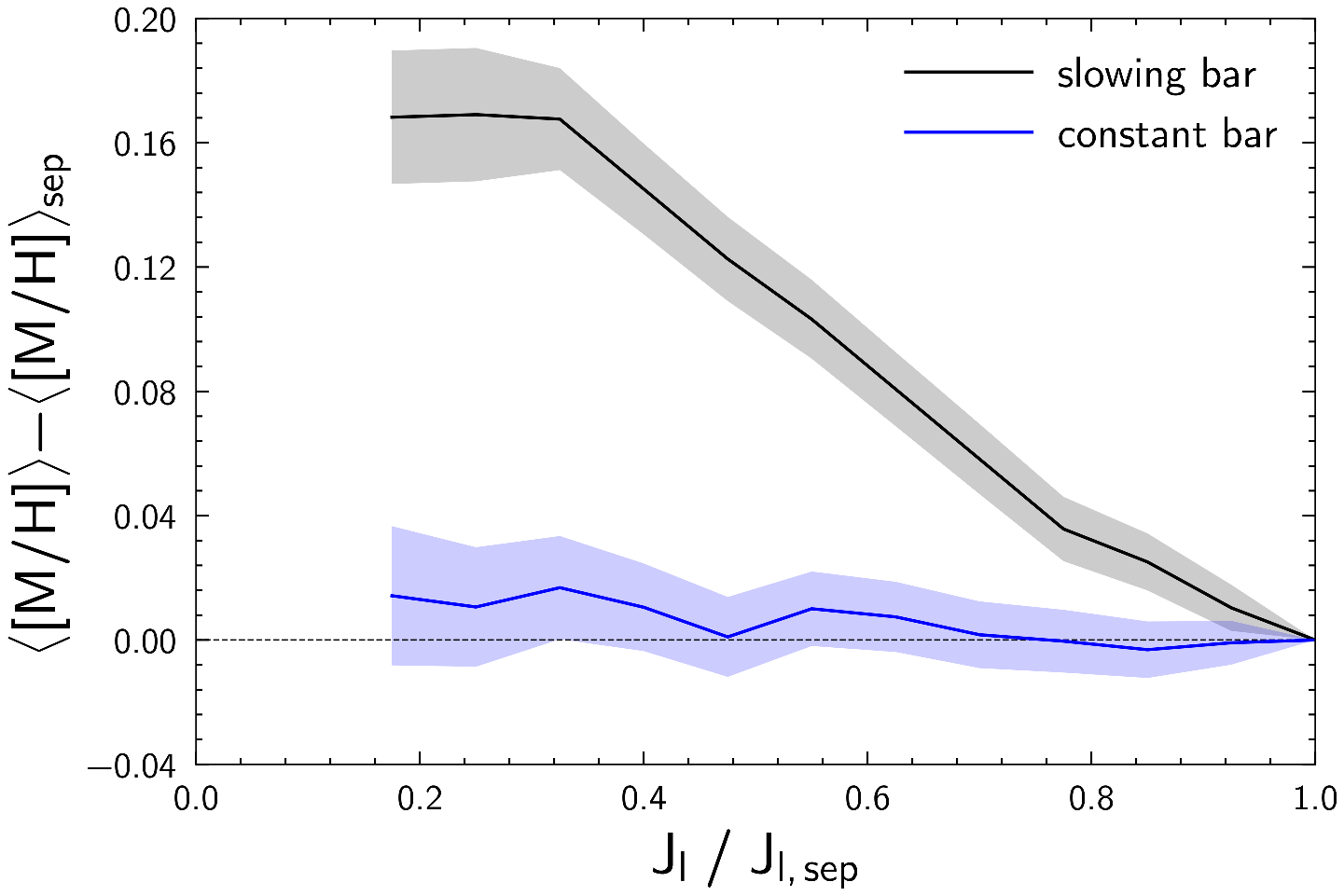}
    \caption{Prediction of mean metallicity $\MH$ inside the resonance using mock data generated from test-particle simulation. Increase in $\MH$ towards small $\Jln$ exclusively happens in the slowing bar model, while there is no appreciable rise in the corresponding model with a constant bar pattern speed.}
    \label{fig:MH_Jl_mock}
  \end{center}
\end{figure}

The bottom panels show the decelerating slow bar model which we have elaborated on in the main text (section \ref{subsec:tree_ring_structure_of_resonance}). The slow down of the bar brings trapped stars from far inside the disc $(\sim 3 \kpc)$ that could potentially have metallicity as high as $0.2 \dex$ in agreement with data. The metallicity-$\Jln$ relation of our models is shown in Fig.~\ref{fig:MH_Jl_mock}. The slowing bar (black) exhibits a profound linear increase in metallicity towards the resonance center up to the initial core ($\Jln \lesssim 0.3$) within which the metallicity is flat as expected. The steady bar (blue) shows no significant rise in metallicity. The result thus corroborates our argument that the observed uptrend in metallicity manifests the deceleration of the bar. We stress though that this is a mere order of magnitude estimation and that many important galactic evolution processes have been ignored, in particular the change in the radial metallicity gradient with time and position as a result of inside-out formation \citep[e.g.][]{Spagna2010,Schoenrich2017Understanding}. Naively, this inside-out signature should flatten or even invert the $\Jphi$-metallicity relationship for old stars (large $\JR$) and thus reduce the metallicity contrast between the CR and the surrounding non-resonant stars at very large $\JR$. A quantitative prediction must await a proper chemo-dynamical model that fully considers these effects.

\bsp	
\label{lastpage}
\end{document}